\documentclass[twocolumn]{svjour3}
\usepackage{mathptmx} 
\usepackage{amsmath}
\usepackage{mdwlist} 
\usepackage{booktabs} 
\usepackage{pgfplotstable} 
\usepackage{textcomp} 
\usepackage{makecell} 
\usepackage[ruled,vlined]{algorithm2e}
\usepackage[caption=false]{subfig}
\usepackage{graphicx}
\usepackage{xcolor}
\usepackage{colortbl} 
\usepackage[normalem]{ulem} 
\usepackage{url}
\usepackage{paralist}
\usepackage{balance}
\usepackage{listings}
\usepackage{verbatim}
\usepackage{array}
\usepackage[flushleft]{threeparttable}
\usepackage{tabularx}
\usepackage{multirow}
\usepackage[shortcuts]{extdash}
\usepackage{etoolbox} 
\usepackage[kerning,protrusion=true,stretch=10,shrink=10,final]{microtype} 

\normalsize

\makeatletter
\renewcommand{\paragraph}{%
  \@startsection{paragraph}{4}%
  {\z@}{1.25ex \@plus 1ex \@minus .2ex}
  {-.1em}
  {\normalfont\normalsize\itshape}%
}
\makeatother

\newcommand{\tick}{TIC}
\def\benchmarksurl{\url{https://github.com/TRDDC-TUM/wcet-benchmarks}}

\AtBeginDocument{%
  \IfFileExists{wasysym.sty}{%
    \RequirePackage{wasysym}%
    \newcommand{\instrumented}{$\CIRCLE$}%
    \newcommand{\sliced}{$\LEFTcircle$}%
    \newcommand{\accelerated}{$\Leftcircle$}%
    \newcommand{\abstracted}{$\Diamond$}%
  }{%
    \newcommand{\instrumented}{instr.}%
    \newcommand{\sliced}{slic.}%
    \newcommand{\accelerated}{accel.}%
    \newcommand{\abstracted}{abstr.}%
  }%
}
\newcommand{\tablelegend}{{\instrumented~instrumented, \sliced~sliced, \accelerated~accelerated, \abstracted~abstracted}}

\newcolumntype{L}[1]{>{\raggedright\arraybackslash}p{#1}}
\newcommand{\redsout}[1]{\color{red}{\sout{#1}}}
\lstdefinestyle{MyListStyle} {
	numbers=left,
	numbersep=3pt,
	language=C,
	basicstyle=\fontsize{6.5}{6.5}\selectfont\ttfamily,
	keepspaces=true,
	breaklines=true,
	breakatwhitespace=false,
	tabsize=2
    }
		
\lstdefinestyle{MyListStyle1} {
	numbersep=3pt,
	language=C,
	basicstyle=\fontsize{7}{7}\selectfont\ttfamily,
	keepspaces=true,
	breaklines=true,
	breakatwhitespace=false,
	tabsize=2
    }

\begin{document}
\title{Scalable and Precise Estimation and Debugging of the\\Worst-Case Execution Time
for Analysis-Friendly Processors}

\subtitle{A Comeback of Model Checking }

\author{Martin Becker\thanks{\email{martin.becker@tum.de}, Tel.+49-89-289-23556}\textsuperscript{1}\and Ravindra Metta\textsuperscript{2}\and R Venkatesh\textsuperscript{2}\and Samarjit Chakraborty\textsuperscript{1}
}                     
%
%
\institute{\textsuperscript{1}Chair of Real-Time Computer Systems, Technical University of Munich, Munich, Germany \and \textsuperscript{2}Tata Research Development and Design Centre, Pune, India}
\date{This is a pre-print of an article published in the \emph{International
  Journal on Software Tools for Technology Transfer}. The final
  authenticated version is available online at:
  \url{https://doi.org/TODO}}
%
%
\authorrunning{Becker et al.}
\maketitle

\begin{abstract}
  Estimating the Worst\-/Case Exe\-cu\-tion Time (WCET) of an application is an
  essential task in the context of developing real-time or safety-critical
  software, but it is also a complex and error-prone process.  
  Conventional approaches require at least some manual inputs from the user, 
  such as loop bounds and infeasible path information, which
  are hard to obtain and can lead to unsafe results if they are
  incorrect. This is aggravated by the lack of a comprehensive
  explanation of the WCET estimate, i.e., a specific trace showing 
  how WCET was reached. It is therefore hard to spot incorrect
  inputs and hard to improve the worst-case timing of the application.
  Meanwhile, modern processors have reached a complexity that refutes analysis
  and puts more and more burden on the practitioner.
  In this article we show how all of these issues can be significantly 
  mitigated or even solved, if we use processors that are amenable to
  WCET analysis. We define and identify such processors, and then we 
  propose an automated tool set which estimates a precise WCET 
  without unsafe manual inputs, 
  and also reconstructs a maximum-detail view of the WCET path that can be 
  examined in a debugger environment.
  Our approach is based on Model Checking, which however is known to scale
  badly with growing application size. We address this issue by shifting the
  analysis to source code level, where source code transformations can
  be applied that retain the timing behavior, but reduce the
  complexity. 
  Our experiments show that fast and precise estimates can be achieved
  with Model Checking, that its scalability can even exceed current
  approaches, and that new opportunities arise in the context of
  ``timing debugging''.
\end{abstract}


\section{Introduction}

Many real-time systems need to provide strict guarantees for response
times. For instance, airbag deployment in a
car, collision avoidance systems in aircraft, and control systems in spacecraft have to meet deadlines for ensuring vehicle and passenger
safety. 
The need to analyze this class of time-critical systems has been the main motivator~\cite{Puschner89} for research on calculating the Worst\-/Case Execution Time (WCET), which is the longest time a
program takes to terminate, considering all possible inputs 
and control flows.

This timing depends both on the structure of the program,
as well as on the processor it is running on.
In the most general case, the WCET problem is undecidable (e.g.,
because loop bounds have to be known), 
and hence only an upper bound
of the WCET can be determined through automatic analysis. Therefore,
in a practical setting, the problem reduces to finding the \emph{tightest} safe upper bound, and in particular
using a technique that scales well with program complexity.

The techniques being applied today for analyzing the timing behavior, 
such as Integer Linear Programming (ILP) and Abstract
Interpretation (AI), or a combination thereof~\cite{WilhelmWCET08},
work very well for analyzing complex programs, but they exhibit
several weaknesses:
\begin{inparaenum}[(1)]
\item User annotations, such as loop bounds, have to be
  provided~\cite{WilhelmWCET08,Souyris2005}, but are hard to obtain,
  influence the tightness and may even refute the soundness of the
  WCET estimate. Providing too large bounds leads to a large
  overestimation, and too small bounds may yield an unsafe
  estimate. As a result, providing safe and tight bounds has become a
  research field on its own with a wide range of different approaches,
  e.g., using Abstract Execution~\cite{Gustafsson2006}, refinement
  invariants~\cite{Gulwani2009} and pattern
  matching~\cite{healy2000supporting}.
%
\item Existing approaches work at the machine code level, where the
  high-level information from the original program is hard to extract.
  Variables are distributed over multiple registers, type information
  is lost, loops and conditional statements can be implemented in many
  different ways, and indirect addressing can make it close to
  impossible to track data flows and function calls. As a consequence,
  overapproximations have to be used, which usually result in pessimistic 
  estimates.
\item Overapproximations are also used to bound the control flow. As a
  result, the control flow that would correspond to WCET might not
  even exist in many cases, contributing further to less tight estimates.
\item Finally, practitioners are facing another challenge with
  today's analysis tools. Once the WCET of an application has been
  computed, the output offers little to no explanation of how the WCET
  has developed, even less of how it can be influenced through changes
  in the program.
\end{inparaenum}
Clearly, there is room for improvement in the process of WCET estimation.

\begin{figure}
  \centering
  \includegraphics[width=.9\columnwidth]{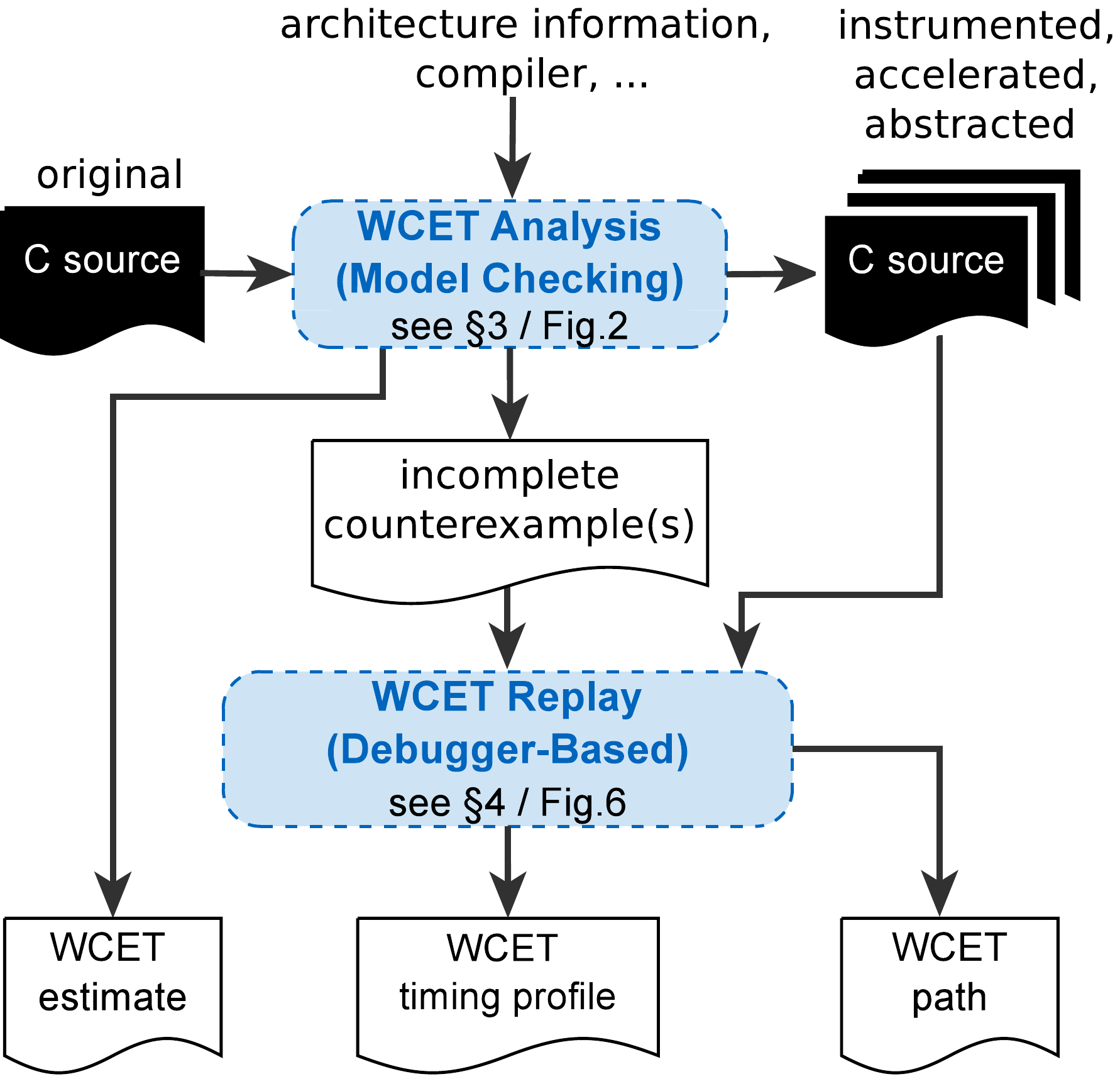}
  \caption{Overview of our WCET tools and their artifacts}
  \label{fig:intro}
\end{figure}

However, there exists an even more fundamental problem: Analysis
cannot keep up with modern processor architecture. In recent years,
the research community arrived at the alarming conclusion that a sound
analysis of modern processors is almost impossible, since their
complex microarchitecture requires considerably large models, which in
turn lead to a state-space explosion~\cite[Chapter 5.20]{mitra_et_al:DR:2017:6953,Puschner2002,PRET2008,DBLP:conf/isorc/KirnerP08}.

In this article, we see this discouraging situation as a chance to anticipate what improvements could be possible in WCET analysis if processors would become more analyzable, as being proposed by various researchers \cite{PRET2008,Puschner2002}.
Assuming that we have a ``deterministic processor'', we question the techniques being used today. 
Specifically, we show that by virtue of more analyzable processors, WCET analysis can build on
 approaches that have already been declared dead for that purpose. 
The weaknesses in the estimation process, such as the difficulty of providing flow annotations and 
imprecision of the estimate, can be significantly reduced, and the usability of WCET tools increased. Moreover, we get a chance to realize a true ``time debugging'', through which practitioners get a handle on the timing behavior, in the same way as functional behavior is being debugged today.


\noindent\textbf{The Idea:} Assuming that we have a processor that is amenable to WCET analysis (we specify and identify such processors later on), our approach builds on the following ideas:

\begin{itemize}
\item{First,} we propose WCET estimation by analysis of the
  \emph{source code}, instead of the machine code. This is enabled by
  choosing analysis-friendly processors, and makes the control flows easier to
  analyze, and the additional information, such as variable types and
  their ranges, helps cutting down the number possibilities to be
  analyzed, thereby providing more precise estimates in shorter time.

\item{Second,} we propose to use \emph{Model Checking} instead of
  traditional ILP or AI techniques for path analysis, because it can precisely
  determine the longest feasible path, as it explores all paths in the
  program explicitly. Therefore, we expect tighter estimates, and ideally 
  we no longer need flow constraints to
  be specified by the user, which makes WCET analysis safer and easier to
  apply.

\item{Last} but not least, we propose to leverage the counterexample
  that is generated by the model checker, to reconstruct the precise
  path that is taken in the WCET case, and enable an \emph{interactive
  replay} that shows not only the path being taken and how time is
  accumulating, but also the contents of all program variables.

\end{itemize}

However, there are three main challenges in realizing our idea: First,
micro\-archi\-tec\-ture\--specific execution times from the
machine-level code need to be represented in the source code. Towards
that, a back-mapping from machine-level code to source code has to be
developed.  
Second, Model Checking had been considered for WCET estimation earlier, but it was
found to scale poorly with program size~\cite{wilhelm-no-mc,Lv2008}, and thus never
been seen as a competitive technique for WCET estimation.
We show that this scalability problem can be mitigated by the
mentioned shift of the analysis from machine code to source code
level, and that source transformation techniques, such as
program slicing and loop summarization, can be used to reduce
the complexity and allow Model Checking to scale better.  The third
challenge is to reconstruct the WCET path, such that it is accessible
to the user, and provides specific explanation why a path was
taken. That is not directly possible from the counterexample, as this
carries incomplete information. We therefore need to devise an
efficient path reconstruction technique.

\noindent\textbf{Summary of our approach:} The overall approach consists of two steps, as depicted in Fig.~\ref{fig:intro}.
First, we estimate the WCET using a model checker, then we reconstruct the WCET path from the counterexample that is provided.
The WCET estimation -- detailed in Fig.~\ref{fig:analysis} -- starts by establishing a mapping between machine code and
source code, and evaluate the instruction timing of the program.
Given this
information, 
we annotate 
the source code with increments to a counter variable, which 
is updated at the end of each source block according to the instruction timing.
The resulting code is then sliced with respect to time
annotations, to remove all computations not affecting the control
flow.  In the next step, we accelerate all loops that can be
accelerated.  Next, we over-approximate the remaining loops with a
large or unknown number of iterations. Then, we perform an iterative
search procedure with a model checker to determine the WCET value. We
terminate the search procedure as soon as we find a WCET estimate
within the precision specified by the user. Finally, the path reconstruction takes place in
a debugger, while forcing decision variables to the values
given in the counterexample, and through that reconstruct the precise
path leading to WCET, whilst collecting a timing profile similar to
\emph{gprof}.


The contributions are as follows:
\begin{itemize}

\item Efficient application of Model Checking at source code level to find the worst\-/case execution time of a C program (Section~\ref{sec:wcet}).

\item Application of source code transformation techniques to reduce the complexity of the system subject to Model Checking (Section~\ref{sec:slicing}ff).


\item A debugger-based technique to reconstruct and replay the precise control flow, all variable contents, and a timing profile of the path leading to the worst\-/case execution time (Section~\ref{sec:replaying-wcet-trace}).

\item A prototype of a tool set called \tick\, which implements our proposed approach.

\item Experiments with the standard M\"alardalen WCET Benchmark Suite, to assess the impact of the source code transformations on scalability and tightness of the WCET estimates, in comparison to an ILP-based analyzer and a cycle-accurate simulator (Section~\ref{expts}).

\end{itemize}  
This article is an extension of our earlier work~\cite{metta2016tic}.




\section{Technical Background}

\label{sec:background}
\subsection{WCET Analysis}
The goal of WCET analysis is to estimate the longest time a
(sub)program $P$ takes to terminate, while considering all
possible inputs and control flows that might occur, but excluding any
waiting times caused by sleep states or interruption by other
processes. In real-time systems, this estimate is subsequently used as
an input for \emph{schedulability analysis}, which then models the
influence of other processes and computes an upper bound of the
\emph{reaction time} of $P$. The reaction time, finally, should be
shorter than any deadline imposed on $P$. For example, the deadline
for $P$ could be given by the maximum time that is permissible to
detect a car crash and activate the airbags.  Consequently, the WCET
estimate is a vital metric for real-time systems, and thus needs to be
\emph{safe} (i.e., never smaller than what can be observed when
executing $P$) and \emph{tight} (i.e., as close as possible to the
observed value).

The WCET of an application is influenced by the processor
architecture, e.g., caches and pipelines, as well as program
structure. Therefore, WCET analysis usually comprises the following
steps (not necessarily in that order):

\begin{enumerate*}
\item \textbf{Compilation:}\label{sec:wcet-analysis-cc} Cross-compile
  $P$ for the processor it is supposed to run on. The source code of
  $P$ is translated to machine instructions $I$, applying various
  optimizations.
\item \textbf{Flow Analysis:}\label{sec:wcet-analysis-flow} Analyze
  $I$ to discover all possible control flows. This includes finding
  all potential branches in $I$ and storing them in a control flow
  graph $G$, including their branch conditions.
\item \textbf{Value Analysis:}\label{sec:wcet-analysis-value}
  Calculate possible ranges for operand values in $I$, to resolve
  indirect jumps and classify memory accesses into different memory
  regions (e.g., slow DRAM vs. fast core-coupled memory).
\item \textbf{Loop Analysis:}\label{sec:wcet-analysis-bound} Bound the
  control flow of $G$, that is, identify loops and compute their
  maximum execution counts based on branch conditions, and annotate
  the nodes and edges in $G$ with those execution counts.
\item \textbf{Microarchitectural
    Analysis:}\label{sec:wcet-analysis-caches} Predict the timing effects of
  caches, pipelines and other architecture-dependent constructs, based
  on memory mapping and the paths in $G$. Annotate nodes and edges in $G$
  with instruction timing considering these features.
\item \textbf{Path Analysis:}\label{sec:wcet-analysis-formulate}
  Formulate a mathematical model based on $G$ and solve for WCET. The
  discovered control flow graph $G$ and the computed loop bounds are
  analyzed together to find those paths along $G$ which could produce
  the longest execution time.
\end{enumerate*}

Steps 2 through 5 are often referred to as \emph{low-level analysis},
and step 6 as \emph{high-level analysis}.  The employed methods
typically involve a combination of Abstract Interpretation and Linear
Programming~\cite{wilhelm-no-mc}: Flow analysis parses the
ISA-specific binary and builds the control flow graph, value and
loop analysis typically use Abstract Interpretation to deduce variable
values and loop counts that may influence control flow or timing,
microarchitectural analysis typically builds on Abstract
Interpretation to approximate cache and pipeline effects, and finally
path analysis is usually done by translating the annotated control flow
graph into a constrained optimization problem~\cite{Li1997}.

\subsubsection{WCET-Amenable Processors}
\label{sec:wcet-processors}
In this paper, we focus on ``predictable'' hardware, as recently
defined by Axner
et al.~\cite{DBLP:journals/tecs/AxerEFGGGJMRRSHW014}. In particular,
the ideal processor for WCET analysis has a \emph{constant}
instruction timing. By this we mean, that each instruction takes a
constant and bounded amount of time, and this time should neither
depend on processor states or operand values, nor be subject to
additional waiting states (e.g., pipeline stalls due to pending bus
transfers). We allow an exception for branch/jump instructions, where
variable instruction timing (e.g., taking a branch may take more time
vs. not taking it) can be covered in our flow analysis. With such a
WCET\-/amenable processor, it is neither necessary to perform a value
analysis at register level, nor do we require complex
microarchitectural models.  Although these requirements seem
unrealistic even for simple processors, they do not
automatically forbid the use of features such as pipelines and caches,
as we shall explain in the following.

\paragraph{Pipeline Requirements.} Processors may have a pipeline, but
the timing of successive instructions must not depend on each
other. Again, we allow one exception for conditional jumps (decisions
upon the control flow), where we model the variable
timing. Furthermore, the processor may make use of bypass/operand
forwarding, but potential bubbles are assumed to be included in the
instruction timing, or otherwise avoided by the compiler;
otherwise, an architectural way around this problem are interleaved
pipelines~\cite{DBLP:conf/iccd/EdwardsKLLPS09}. Furthermore, there
must be no out-of-order processing. Through that, structural and data
hazards need not be modeled, but only control hazards at the
granularity of basic blocks.

\paragraph{Cache Requirements.} Processors may have instruction and
data caches, but we do not allow for cache misses. That is, cache
contents are selected at or prior to compile time and kept static
during execution, or deterministically loaded by software at locations
known at compile time. For example, caches could be loaded every time
a function is entered (as in \cite{JOP}), or explicitly through
statements in the source code (scratchpad memory). Cache
locking~\cite{Mittal16} provides an alternative mechanism to reach the
same effect (which, incidentally, can improve the
WCET~\cite{Ding2004}), and potentially qualifies many more processors
for our approach. With these cache requirements, timing effects due to
caches can be annotated in the source code as part of our
analysis, and do not have to be modeled on instruction granularity.

\paragraph{On-Chip Bus Transfers.}
In many system-on-chip processors, there are peripherals (e.g,. a UART
peripheral) which are accessed from the core via on-chip buses.  Since
usually there will be no model available for the behavior of the
peripheral, we cannot support such accesses in a WCET analysis. A
WCET\-/amenable processor therefore should provide time bounds for 
instructions performing such accesses. Consequently, no model is
necessary for bus arbiters and peripheral states. 

\paragraph{Multicore Processors.}
With the presence of hardware threads, there could be interference
caused by resource sharing. In principle, there are techniques which
provide some temporal isolation for the hardware threads, as described
in \cite{DBLP:journals/tecs/AxerEFGGGJMRRSHW014}. We expect 
that a WCET\-/amenable processor has to implement such techniques, since
otherwise the WCET problem becomes even more intractable than it
already is for today's monoprocessors. Consequently, we are only
considering mo\-no\-pro\-cess\-ors here, assuming that an extension to
multicore processors can build on such isolation, and otherwise
capture high-level interactions such that they are visible in the
source code.

\paragraph{Amenable Processors.} Due to the complexity and variety of
modern processor architectures, only a detailed review on a
case-by-case basis allows to identify existing processors which
fulfill our requirements. Therefore, we can only give a few select
examples, and otherwise refer to our requirements for what we wish
future processor architectures would look like, to keep WCET analysis
tractable.  Slight overapproximations, such as using only the maximum
execution time for time-variable instruction, might be applied to use
our approach even for processors that are not strictly compliant.  The
processor family that we target in this paper, the 8-bit Atmel AVR
family~\cite{AVRdatabook}, is a good example for that. While there is a dependency of the
instruction timing on the operands in some cases -- namely, slight
variations in instruction timing for flash memory access -- this is
negligible and can be overapproximated. Our results shown later in
this article justify this approach. Other processors that are a good
fit are the SPARC V7 (specifically the ERC32 model~\cite{SPARCv7}), the ARM7TDMI
family~\cite{ARMv7TDMI}, and the Analog Devices ADSP-21020. Academic examples include
the Java-Optimized Processor~\cite{JOP}, and the Precision Timed
Architecture~\cite{PRET2008} with minor modifications (namely,
port-based I/O to avoid time variances in load/store instructions, and
absence of structural hazards).


\subsubsection{WCET Analysis at Source Code Level}
Our main goal is to shift WCET analysis from the instruction level to
the source code level, where control flows are easier to follow and
type information is available. We expect two profound effects from
this shift. First, a model checker can leverage the clearly visible
control flow and type information to perform an automatic path
analysis, without requiring user inputs. That should lead to tighter
and safer estimates than other approaches, such as ILP. Second, the
complexity of the analysis is reduced, since operations are
represented in a more high-level view.

A necessary prerequisite for such a source-level analysis is to
annotate the source code with the instruction timing, as faithfully as
possible.  We do so by introducing a counter
variable~\cite{Holsti2008} into the source code, which is a global
variable representing execution time that is incremented after each
statement according to the time taken thereof (see Fig.~\ref{fig:ins}
for an example).

Towards that, timing information needs to be extracted from the
executable code, whilst considering the microarchitecture of the
target processor, and then mapped back to the source code in the form
of assignments to our counter variable at the correct locations. For a
WCET\-/amenable processor, considering the microarchitecture boils down
to model variable instruction timing only at branch points. The time
annotations in the source code must therefore allow the encoding of such
variable timing. In principle this is only possible precisely when the control
flow of the instructions is also mapped back to the source code. That
is, conditional jumps in the instructions must be lifted in the source
code.  Later, we show that some overapproximations can be applied, in
case a back-mapping is not possible because a source equivalent is lacking
for some instructions.

In general, mapping the instruction timing back to the source code
means to establish a mapping from the instructions to source-level
constructs.  This can be difficult, because compiler optimization may
produce a control flow in the executable that is very different from
that of the source.
For example, functions that are present at the source level may be
absent in the executable because of inlining. Vice versa, new
functions could be introduced in the executable which have no direct
match in the source code.  For instance, it is common for compilers to
introduce loops and functions to compute 64-bit multiplications and
shifts, or to implement switch case statements as binary search or
lookup tables. These kinds of transformations make it hard to
automatically map the basic blocks in the machine instructions to the
source code.  An ideal compiler would keep track of the mapping during
the translation process, however, we are not aware of any compiler
doing so. Therefore, some over-approximations must be applied to
overcome these difficulties, and they can go a long way in automating
the back mapping.  

For the work in this paper, we have established a heuristic mapping
for the AVR gcc compiler with only little difficulty, as explained
later in Section~\ref{sec:back-annotate}. Other researchers have
accomplished the same goal for different WCET\-/amenable
processors~\cite{Kim2009,Raymond2013} even under some optimization,
suggesting that such a mapping usually can be established when only
considering the timing.

In the next section, we elaborate how Model Checking can be used on
 this time-annotated 
source code to estimate the WCET.

\subsection{Model Checking\label{Background:CBMC}}
Model Checking~\cite{Clarke:2000} is a formal analysis technique used
to verify a property on a model. Given a model and a property, a
\emph{model checker} -- a tool implementing this technique --
determines if the model satisfies the property in every possible
execution. If the property does not hold, the model checker produces
evidence for the violation, called a counterexample.  Model checkers
perform reachability analysis over finite-state transition systems,
where the number of states is a product of program locations and
variable valuations at these locations. Therefore, though Model
Checking is sound and complete for finite-state models, scalability is
often an issue for complex models.

It has been demonstrated that model checkers are useful at computing
the WCET of simple programs~\cite{Kim2009,Kuo2010}, but the
scalability issue has not been addressed before.  The idea is to take
a program that is annotated with timing information, translate it into
a model and use a model checker to verify the property ``at program
exit, execution time is always less than $X$'', where $X$ is a
proposed WCET value. The model checker acts as an oracle, telling if
there exists any execution that takes longer than this proposal. This
process is repeated with changing proposals, until we find the lowest
upper bound where no counterexample is generated. We will follow the
same approach here, but address the scalability issue by minimizing
the number of oracle queries, and also the complexity of the
individual queries.

For our experiments we have used the model checker CBMC~\cite{Clarke2004}, due to its robustness and performance. It is a \emph{bounded model checker} which accepts models and properties in the form of ANSI-C programs. 
Some important features of CBMC that we use in WCET computation are:
\vspace{-.25em}
\begin{itemize}
\item \textbf{Assertions}: CBMC allows expressing properties with 
  assertions. In our case, these assertions are of the form {\tt
    assert(\_time$<$X)}, where the constant {\tt X} is the proposed WCET,
  and {\tt \_time} denotes the counter variable reflecting the time passing by in the program.

\item \textbf{Non-determinism}: CBMC allows any of the program variables to be assigned
a non\-/deterministic value. This is done using
assignments of the form {\tt y$=$nondet()}, where {\tt nondet()} is an undefined function having the same return type as {\tt y}. 
This results in \texttt{y} being assigned a non\-/deterministic value from the range specified by its type.

\item \textbf{Assumptions}: CBMC allows the use of {\tt assume} statements to
  block the analysis of undesirable paths in a program. A statement of
  the form {\tt assume(c)} marks all those paths infeasible for which
  {\tt c} evaluates to {\tt false} at the execution of this
  statement. This feature can be used to constrain the value domain
  of non\-/deterministic assignments.

\item \textbf{Checking multiple assertions}: CBMC allows multiple assertions in
  the input program, which can be checked at once using the {\tt
    --all-properties} 
  option.  This option uses an optimal number of solver calls to
  verify programs with multiple assertions.

\end{itemize}

The technique we present here does not depend on CBMC specifically. It could be replaced by any other model checker, possibly through an additional front-end that translates C code into a model to work with (e.g., the CPROVER tools).

\subsection{Program Slicing\label{background:slicing}}

\begin{figure*}[tb]
  \includegraphics[width=.95\textwidth]{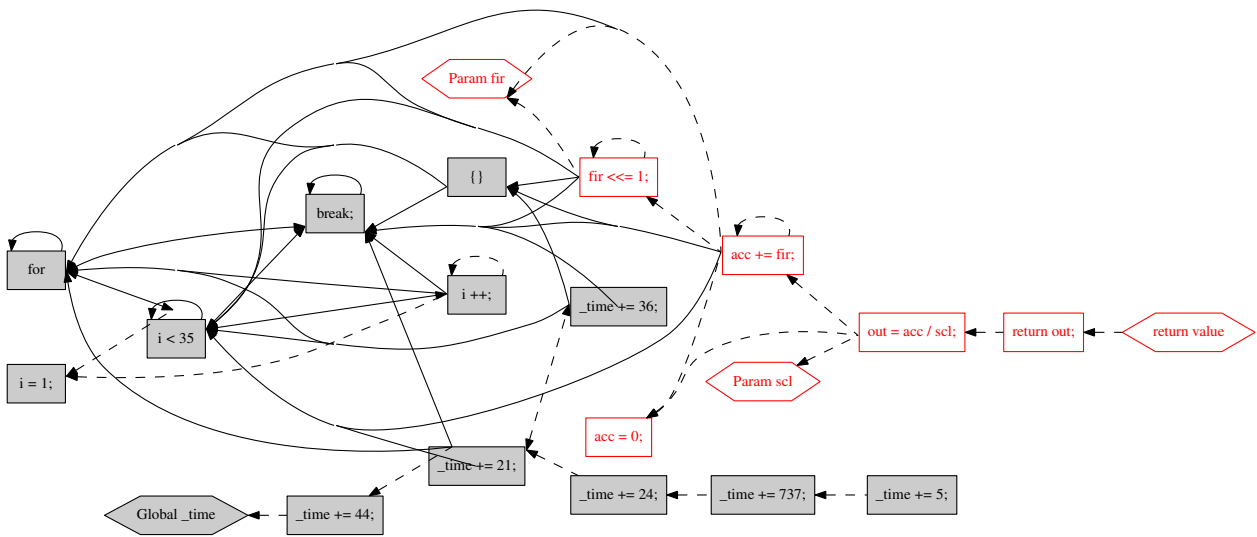}
  \caption{Program Dependence Graph of the program from Fig.~\ref{fig:sli}. Sliced statements are outlined/red}
   \label{fig:slicing}
\end{figure*}

Program slicing was first introduced by Mark Weiser in
1981~\cite{Weiser:1981}.  Given an imperative program and a slicing
criterion, a program slicer uses data flow and control flow analysis
to eliminate those parts of the program that do not impact the slicing
criterion.  That is, it removes statements which do not influence the
control flow related to the criterion, and do not change its value.
The resulting program is called a ``program slice'', and behaves
identically w.r.t. the slicing criterion, but has a reduced
complexity.

Slicing works by constructing and evaluating a Program Dependency
Graph (PDG), which captures the control and data flow dependencies of a
program. For example, Figure~\ref{fig:slicing} shows the PDG
corresponding to the code in Figure~\ref{fig:ins}. This graph has
two kinds of edges to denote dependencies: dashed edges denote
data dependence, and solid edges denote a control flow
dependency. For example, the loop increment ``i++'' is
control-dependent on the loop condition ``i $<$ 35'', and 
data-dependent on itself as well as on the loop initialization.

Given the PDG from Fig.~\ref{fig:slicing} and a slicing criterion, we start at the node that
corresponds to the location of the criterion, and traverse the PDG
from this point until all the root nodes of the graph are reached.
Subsequently, we remove from the program all statements and
expressions that correspond to nodes that have not been visited
during this traversal. For our example program in Fig.~\ref{fig:ins},
the criterion is the latest possible assignment to variable
\texttt{\_time} in line~17, and the corresponding PDG node is ``\_time
+= 5'' (in the lower right corner of the graph). When
traversing the graph from there, the outlined/red nodes (e.g., ``out =
acc/scl'') are not reachable. Therefore, these parts of the program do
not impact the value of variable \texttt{\_time} at our location, and
can be safely removed from the program. The resulting program slice is
given in Fig.~\ref{fig:sli}.

\subsection{Loop Acceleration}
\label{background:accel}
\emph{Loop acceleration} describes the action of replacing a loop with a precise closed-form formula
capturing the effects of the loop upon its termination. This has been shown to be effective for
Model Checking of source code ~\cite{labmc,SVCOMP}.

For loop acceleration to be applicable, the loop should have the following characteristics:
\begin{itemize}
\item The loop should iterate a fixed number of times, say $n$, which
  could either be a constant or a symbolic expression.  For example,
  the loop may execute 10 times or $n$ times, or $x*y$ times and so
  on.
\item The statements in the loop constitute only of assignments and
  linear arithmetic operators, that is, first order polynomial
  computations (and no higher order).
\item The loop body consist of straight line code, that is, there are
  no branching statements such as if-else statements inside the loop
  body.
\end{itemize}

When a loop satisfies the above constraints, it is possible to replace
the loop with simple linear recurrence relations that precisely
compute the summarized effect of all the iterations of loop on each of
the assignments in the loop body. For example, the for-loop in
Figure~\ref{fig:sli} is replaced with the block spanning lines 7
through 13 in Figure~\ref{fig:acc}.


\subsection{Abstraction}
\label{background:abs}
\emph{Abstraction} is a term used to describe a
modification to an input program in such a way that the resulting
output program allows more runs than the input program. That is, the
set of all possible program states at the end of the output program,
is a superset of all possible states at the end of the input
program. In other words, abstraction creates a safe overapproximation
of a program. 

For instance, in Section~\ref{background:accel}, suppose a loop has
only the first two characteristics, but violates the last one,
viz., the loop does have branching statements in the form of if-else
statements.  Then, it is still possible to replace the loop with an
over-approximate formula. The most simple way to achieve this is to
replace it with a non\-/deterministic assignment that allows for the
computation of the result in any arbitrary value. For example, the
effect of the loop in Figure~\ref{fig:ex2} on the variable
\texttt{ans} is abstracted as shown in line~5 of
Figure~\ref{fig:abs}. There we allow \texttt{ans} to take on any
non\-/deterministic value in its type range, which is a superset of all
the feasible values among all possible executions.



\section{Finding the Worst\-/Case Execution Time}

\begin{figure}
  \centering
  \includegraphics[width=\columnwidth]{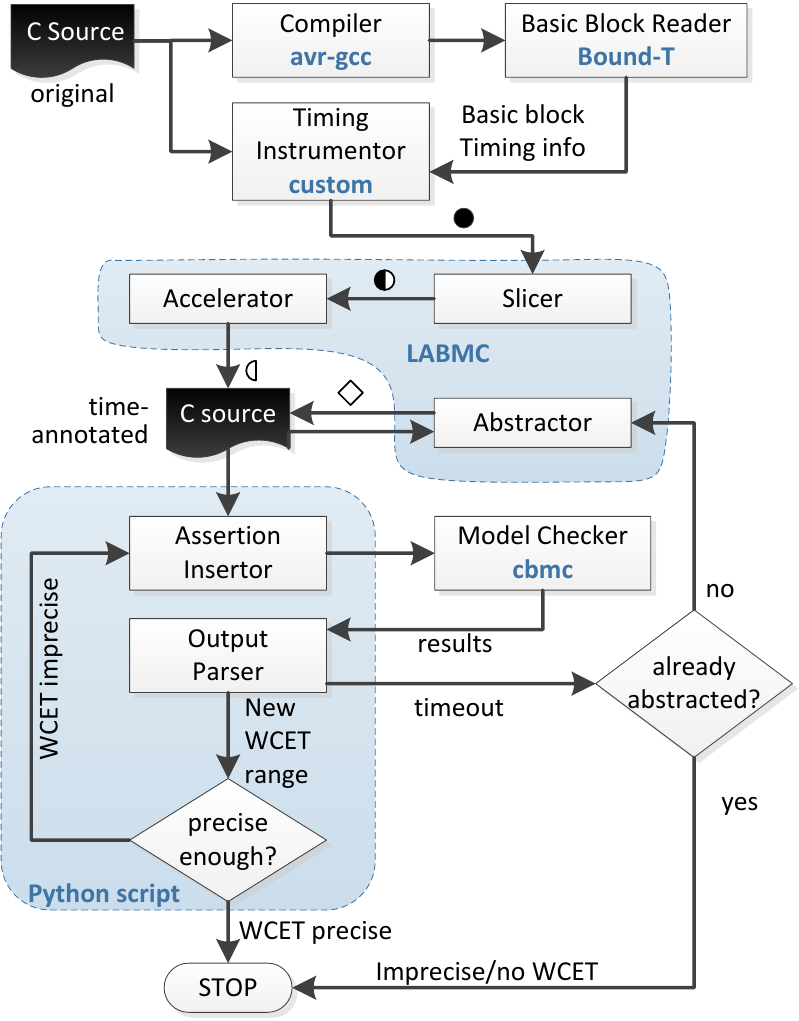}
  \caption{WCET analysis using Model Checking}
  \label{fig:analysis}
\end{figure}

Our tool set \tick\ estimates the WCET at source level, with the help
of timing information that is extracted from the executable code. The overall 
workflow is illustrated in Fig.~\ref{fig:analysis} and shall now be explained in detail.
Given a C program, a corresponding executable that results from cross-compilation for
the target, and information about the target architecture, \tick\ does the following:
\begin{enumerate}
\item Estimates the time taken for each basic block in the machine code of the program. 
\item Establishes a mapping from these basic blocks to a set of source lines and instruments the C program with the timing obtained in the previous steps. 
\item Detects and prepares the handling of persistent variables, i.e., the internal states of the program, since those may have an impact on the control flow and thus on the WCET. 
\item Applies source code transformations to eliminate all irrelevant data computations and to summarize loops. 
\item Adds a driver function which represents the entry point for the model checker and encodes the search for WCET as a property. 

\item Uses a model checker and an appropriate search strategy to estimate the WCET with desired precision. 
\item If the model checker does not scale, then it abstracts the program and repeats the search.
\end{enumerate}

The details of each of the above steps are given in the following
sections. As a running example, we use a simplified version of the 
\emph{fir} benchmark from the M\"alardalen WCET benchmark suite.

\subsection{Estimation of Basic Block Time (\instrumented)\label{sec:step-1:-backtr}}
We first analyze the executable file produced by the target-specific
compiler to construct a control flow graph. Towards that, we identify
\emph{basic blocks}. These are maximal sequences of instructions that
are free of branches, i.e., where the only entry point is the first
instruction and the only exit point is the last instruction.
Therefore, the basic blocks become the nodes in our control flow
graph, and the edges represent branches or function calls, labelled
with their branching condition. Finally, we determine the execution
time for each basic block by adding up the time taken by the contained
instructions, and annotate 
our nodes with this block timing. Since we allow for branch
instructions to have a variable timing (e.g., the instruction may take
longer if a conditional branch is taken), we annotate the precise
timing of the jump instructions to the edges of the graph. Through
that, there is no overestimation due to such timing variations.

Our implementation re-uses parts of the Bound-T WCET
analyzer~\cite{holsti2002status} to build the control flow graph and
estimate basic block times. We therefore have exactly the same inputs
for our approach and the ILP-based analysis in Bound-T. Alternatively,
a more generic binary analysis library could be used, such as the
\emph{Binary Analysis Platform}~\cite{brumley2011bap}.

\subsection{Back-Mapping and Annotation of Timing}\label{sec:back-annotate}

\begin{figure*}
  \hspace*{2mm}
  \resizebox{0.97\textwidth}{!}{
    \begin{tabular}[t]{@{}lccr}
      \subfloat[Original code\label{fig:ex1}]{%
\begin{lstlisting}[style=MyListStyle]
int task(int fir, int scl) {
  int i, out, acc=0;
  for(i = 1 ; i < 35; i++) {
      acc += fir;
      fir <<=1;
  }
  out = acc/scl;










  return out;
}
\end{lstlisting}\label{fig:orig}} &
      \subfloat[Instrumented]{%
\begin{lstlisting}[style=MyListStyle,escapechar=@]
#define TIC(t) (_time += (t))
unsigned long _time = 0;

int task(int fir, int scl) {
  int i, out, acc=0;
  TIC(44);   @{\color{gray}{// BB1}}@
  for(i = 1 ;
      TIC(21),   @{\color{gray}{// BB2}}@
      i < 35; i++) {
      acc += fir;
      fir <<=1;
      TIC(36);   @{\color{gray}{// BB3}}@
  }
  TIC(24);   @{\color{gray}{// BB4}}@
  out = acc/scl;
  TIC(737);  @{\color{gray}{// mult. \_divmodsi4}}@
  TIC(5);    @{\color{gray}{// BB6}}@
  return out;
}
\end{lstlisting}\label{fig:ins}} & 
      \subfloat[Sliced]{%
\begin{lstlisting}[style=MyListStyle,escapechar=@]
#define TIC(t) (_time += (t))
unsigned long _time = 0;

@\redsout{int}@ task(@\redsout{int fir, int scl}@) {
  int i, @\redsout{out, acc=0}@;
  TIC(44);
  for(i = 1 ;
      TIC(21),
      i < 35; i++) {
      @\textbf{\redsout{acc += fir;}}@
      @\textbf{\redsout{fir <<= 1;}}@
      TIC(36);
  }
  TIC(24);
  @\textbf{\redsout{ out = acc/scl; }}@
  TIC(737);
  TIC(5);
  @\redsout{return out;}@
}
\end{lstlisting}\label{fig:sli}} & 
      \subfloat[Accelerated]{%
\begin{lstlisting}[mathescape,style=MyListStyle,escapechar=@]
#define TIC(t) (_time += (t))
unsigned long _time = 0;

void task(void) {
  int i, _k0, _k1;
  TIC(44);
  {
    _k0 = 1;
    i = 35;
    _k1 = i - _k0;
    TIC(21*_k1 + 36*_k1);
    TIC(21);
  }
  TIC(24);
  
  TIC(737);
  TIC(5);

}
\end{lstlisting}\label{fig:acc}}%
    \end{tabular}    
  }
  \caption{Example for source code instrumentation and transformations to compute the WCET of a function \texttt{task}}
  \label{fig:fig1}
\end{figure*}


The next task is to match the control flow of the machine instructions with the control flow of
the source code, and back-annotate the basic block timing in the
source code, in the form of increments to a counter variable. The
result shall be a source code that is instrumented with timing,
in the following called \emph{instrumented program}. 

Specifically, here we annotate the basic blocks of the source with the
instruction timing as close as possible to the corresponding source
construct.  That is, each maximal sequence of statements that is free
of branches (just like basic blocks at machine code level) is
immediately followed (or preceded, in the case of loop headers) by a
timing annotation that reflects the execution time of the instructions
corresponding to that block.

First of all, an approximate and incomplete mapping of machine
instructions to source line numbers is given by GNU's \emph{addr2line}
tool. However, it is incomplete because some instructions do not have
a direct match in the source code, and it is only approximate because
it does not provide column numbers. Complex expressions falling into
multiple basic blocks, such as loop headers or calls in compound
expressions, cannot be resolved with this information. We therefore
use this information as an initial mapping, and then apply safe
heuristics to complete the picture. 

For the heuristics that manage the back-annotation of the basic block (BB) timing information
according to source code, there are two cases to handle:
\begin{enumerate}
\item \emph{One-to-Many mapping.}
One basic block in the executable corresponds to one or more continuous expressions in the source code.  
This case is trivial to handle,  all that needs to be done is instrument the basic block before the last expression with the corresponding timing information.

\item \emph{Many-to-one mapping.}
In these cases, several basic blocks in the executable map to a single source expression. 
Typically, this case arises when a source expression splits into different control paths in the machine instructions, 
for example in case of a division or a bit shift operation being converted into loops. 
In such cases, it is hard to instrument the source code with this information, since it would require translating back the loop into the source first. 
To tackle this issue, we summarize the timing information from such multi-path instruction blocks, and instrument the source code with its worst-case timing value. 
\end{enumerate}

\noindent As an example, the timing of the source code in Figure~\ref{fig:orig} was mapped as follows:
\begin{table}[h]
\vspace*{-2em}
  \centering
  \caption{Mapping of basic blocks (BB) to souce code}
    \begin{tabular}{ c c c r l }
      \toprule
      Start line & End line & BB (\#) & Time & Comment \\ 
      \midrule
      1 & 3 & 1 & 44 & including for init \\  
      3 & 3 & 2 & 21 & for conditional \\
      3 & 6 & 3 & 36 & including for iter \\
      7 & 7 & 4 & 24 & before div \\
      7 & 7 & mult. & 737 & div block  \\
      7 & 18 & 6 & 5 & after div  \\
      \bottomrule
    \end{tabular}
\vspace*{-1em}
\end{table}

The table contains timing information of each basic block along with the span of the block in the source code.
This information is used to instrument 
the source code as shown in Figure~\ref{fig:ins}: We introduce a global counter variable \texttt{\_time}, which is incremented through macro \texttt{TIC} by the corresponding amount of time at the respective source locations.

Basic block 2 is an instance of a one-to-many mapping.  This block
maps to the conditional part of the for-statement on line~3 of Figure
\ref{fig:orig}, and therefore the annotation 
is inserted just before the statement on line~8 in
Figure~\ref{fig:ins}.  Similarly, basic blocks 1 and 3 are one-to-many
mappings.  Basic block 1 implements the start of the function line~1,
as well as the declarations and initialization on line~2 and also the
for-initialization block on line~3. All these source level lines fall
into a single source basic
block. 
Similarly, the for-loop increment on line~3 
along with statements on lines~4 and 5 map to basic block~3. The instrumentation 
in this case is placed in lines~6 and 12 respectively, in Fig.~\ref{fig:ins}.

The division assignment in line~7 is an instance of a many-to-one
mapping, i.e., many basic blocks, and therefore also some conditional
jumps, map to this single statement. Here, these basic blocks include
a compiler-inserted function call (for the long division). In
particular, the compiler-inserted function contains a loop, for which
we do not compute the timing precisely, but instead we
over-approximate this loop with its worst-case bound, and subsequently
we use a single timing annotation for the entire function. Thereafter,
the statement at line~7 can be represented by three parts; one before
the division (function call), one for division (function WCET) and the
final one after the call. The resulting three timing annotations are
shown on lines 14, 16 and 17 Figure~\ref{fig:ins}.


This concludes the shift of WCET analysis from machine code level to
source level. At this point, we have an \emph{instrumented program},
i.e., a source code carrying the execution timing, ready to be
analyzed for WCET. From now on, we continue the analysis with the
instrumented source code only.

\subsection{Handling of Persistent Variables\label{sec:handl-pers-vari}}
A WCET estimation of a function $f$ (including all its direct and
indirect callees) must consider all possible inputs to $f$, and from
those derive the longest feasible paths. Such inputs can be function
parameters, but also referenced variables that are persistent between
successive calls of $f$ (they can also be seen as the hidden program
state). In the C language, such persistent variables are 
\emph{static}  and
\emph{global} variables that are referenced by $f$ or its callees.

In this work we over-approximate the content of such persistent variables by
initializing them with a non-\-deterministic value, as explained in
Section~\ref{Background:CBMC}. This guarantees that all the feasible and infeasible 
values of the persistent variables, as allowed by their data type size, are considered as inputs to $f$. Thus, the WCET estimate for $f$ is always a safe over-approximation of the maximum
observable execution time of $f$. It is possible to remove (some of) the infeasible values either with manual inputs by users or by analyzing the callees of $f$. This may lead to a tighter WCET, but we did not explore these approaches as they are orthogonal to our main work.

\subsection{Source Transformation Techniques\label{sec:source-transf-techn}}
The source code is now instrumented with its timing behavior, and
ready to be analyzed for WCET. However, a direct application of Model
Checking to compute the WCET of large and complex programs would not
scale due to the size of the generated model. The analysis time could
quickly reach several hours for seemingly small programs, and memory
requirements may also quickly exceed the available resources. Our next
step, therefore, are source code transformations which retain the
timing behavior, but reduce the program complexity. This can be done
effectively thanks to the additional information available in the
source code, such as data types and clearly visible control flows.
The transformations are executed sequentially in \emph{stages}, which
we explain in the following. All three states require only little
computational effort themselves, and therefore speed up the overall
process of WCET estimation.

\subsubsection{Stage 1: Slicing (\sliced)}\label{sec:slicing}
Slicing reduces the size of the program by removing all
statements that do not influence a certain criterion, as explained in
Section~\ref{background:slicing}. For the specific case of WCET
analysis using a counter variable, our slicing criterion is the value
of this variable upon program termination,
through which statements not impacting the counter variable are
eliminated, and WCET estimation becomes less complex.

As an example, consider again the instrumented program in Figure~\ref{fig:ins}.
Firstly, observe that line~15 is not needed to compute timing, since the variable \texttt{\_time} has no
data or control dependency on the variable \texttt{out}. 
Similarly, lines 10 and 11 do not impact timing computation and can be sliced away.
 The sliced source for this example is shown in Figure~\ref{fig:sli}. 

We used a program slicer that builds an inter-procedural program dependency graph~\cite{Pingali1995}, capturing both intra\-/procedural and inter\-/procedural data and control dependencies. It then runs one pass over this graph to identify statements that impact the property to be verified and outputs the sliced code.
The slicer is a conservative one, which means that it discards statements only when 
 sure that the statements do not affect the timing analysis.  In all other cases,
the statements are preserved.


\subsubsection{Stage 2: Loop Acceleration (\accelerated)\label{sec:acc}}
A major scalability obstacle in Model Checking are loop-like
constructs, since they have to be unrolled before
analysis. They can therefore increase the program complexity
significantly. This problem can be solved by applying loop
acceleration, as described in Section~\ref{background:accel}. The
resulting program will have a reduced number of loops, and therefore
exhibit a reduced complexity.

As an example, consider the loop in Figure~\ref{fig:sli}: Here, \texttt{\_time}
is incremented in line~12 within a loop body through \texttt{TIC}. The effect
of this repeated increment of  \texttt{\_time}, can be summarized 
by the expression \texttt{\_time} = \texttt{\_time} + $n\cdot 36$, where $n$ is
the number of loop iterations.  Line~11 in Figure~\ref{fig:acc} shows the
accelerated assignment. Note that two new variables
\texttt{\_k0} and \texttt{\_k1} have been introduced, representing the initial
value of the loop counter (\texttt{\_k0}) and the number of loop iterations
(\texttt{\_k1}).  After accelerating all variables, the loop in
Figure~\ref{fig:sli} can be removed and replaced by the statements given in
Figure~\ref{fig:acc} lines 7 through 13. Here, lines 8 to 11 represent the effect
of all the 34 iterations of the loop. Line~12 captures the time taken for
evaluating the loop condition after the final iteration.

As the accelerated program of Fig.~\ref{fig:acc} is free of (most)
loops, it is less complex than the instrumented program. For this
example, the program size (as determined by CBMC) reduces from 325
steps for the instrumented program to only 60 steps for the
accelerated program.
Last but not least, note that a prior slicing is important for loop acceleration; if we had not
sliced the instrumented program w.r.t. \texttt{\_time}, then the above loop
could not have been replaced, as the assignment to \texttt{acc} on line~4 of
cannot be accelerated due to the assignment to \texttt{fir}
on line~5.

\tick\ implements acceleration as proposed in~\cite{labmc}, mainly as
they are shown to be effective on SVCOMP benchmark C
programs and industrial programs~\cite{SVCOMP}.

\subsubsection{Stage 3: Loop Abstraction (\abstracted)}\label{sec:abs}
Finally, further reduction of complexity at the cost of precision can
be accomplished by abstraction, as described in
Section~\ref{background:abs}.
However, we tailor the abstractions by including some specific
time-approximations that preserve the WCET estimate and reduce
complexity even further.

\begin{figure}
\hspace{.1cm}
\begin{tabular}{ll}
\subfloat[Branch inside loop\label{fig:ex2}]{
\begin{lstlisting}[style=MyListStyle]
for (i=1 ;
     TIC(20), i<=len ;
     TIC(8), i++) {
   TIC(35);
   if(ans & 0x8000) {
      TIC(31);
      ans ^= 4129;
   } else {
      TIC(23);
      ans <<= 1;
   }
}
\end{lstlisting}
} &
\subfloat[Abstracted loop\label{fig:abs}]{%
\begin{lstlisting}[mathescape,style=MyListStyle,escapechar=@]
int _k0, _k1;
_k0 = 1;
i = len + 1;
_k1 = i - _k0;
ans = nondet();
TIC(_k1*(20+8+35)+_k1*31);
TIC(20);




$~$
\end{lstlisting}%
}\\  
\end{tabular}
\caption{Loop abstraction using LABMC}
\label{fig:fig2}
\end{figure}

Abstraction is used to get rid of loops that could not be
accelerated, such as the loop shown in Figure~\ref{fig:ex2}: The
if-condition on line~5 depends on the variable \texttt{ans}, which is
updated both in the \emph{then}-branch (line~7) and \emph{else}-branch
(line~10). These assignments determine how many times the \emph{then}
and \emph{else}-branches would be executed.  As these branches depend
on values updated within the loop, we cannot determine the number of
times this branch would be executed, and hence we cannot accelerate
the loop.


The abstracted version of the loop in Figure~\ref{fig:ex2} is shown in
Figure~\ref{fig:abs}. First, we introduce variables \texttt{\_k0} and
\texttt{\_k1} to capture the initial value of the loop counter (line~2)
and the value of it upon loop termination (line~4). Then, the variable
\texttt{ans} is assigned a non\-/deterministic value (line~5). This
assignment, as explained in Section~\ref{Background:CBMC}, allows
\texttt{ans} to take any value in the entire range of its data type,
\texttt{int}. While in the original program \texttt{ans} may take only
a subset of \texttt{int} values, we now allow it to take on any
\texttt{int} value, and thus have constructed an abstraction.

After this abstraction, the loop can now be accelerated as explained
before. Line~6 contains the accelerated assignment to
\texttt{\_time}. In this, \texttt{\_k1*(20+8+35)} accounts for the
time increments in Figure~\ref{fig:ex2} corresponding to the loop
condition evaluation (line~2), loop counter increment (line~3) and the
first part of the loop body on line~4. Finally, line~7 captures the
time taken for evaluating the loop condition after the final
iteration.

\paragraph{Time Approximations.}
The trailing \texttt{\_k1*31} in Fig.~\ref{fig:abs} on line~6
summarizes the time taken by the if-statement forming the remainder of
the loop body, but it is somewhat special.  It contains a
WCET-specific modification to the abstraction as explained in
Section~\ref{background:abs}: Since now the values of \texttt{ans}
have been abstracted, we can no longer reason about which branch of
the if-statement is taken in the WCET case. Therefore, when
abstracting such undecidable conditional statements, we over-approximate
the effects on our time variable by only considering the longest
branch (here: the then-branch in line~6 with 31 clock cycles). Note
that this does not change the WCET estimate, because the model checker
would have picked the longer branch anyway, since the possible values
of \texttt{ans} include the case allowing the then-branch to be
taken. It is thus a safe modification to the abstraction in the
context of WCET analysis, which however reduces the complexity of the
program further.  \vspace{1em}

If a loop contains unstructured code, such as \texttt{break} and
\texttt{return} statements, these are easily handled through a
non\-/deterministic Boolean variable that allows these statements to be
executed or not executed.

To implement abstraction, we again used the LABMC
tool~\cite{labmc}. However, the standard abstraction has been tailored
for the counter variable \texttt{\_time}, as explained above, to pick
the maximum time increment from the branches.

\subsubsection{Further transformations}
A number of other source transformations could be applied to further
reduce the program complexity and yet retain the worst-case timing
information.  For example, after the back-annotation of timing, the
counter variable could be summarized by merging increments belonging
to the same source block. Or, live variable analysis and loop folding
could be used to reduce the unwinding depth of a program
\cite{SAFECOMP15}. However, all of this would make it harder to
understand how source-level statements contribute to timing, and thus
has not been investigated in the context of this work.

Furthermore, CBMC itself (more specifically, goto\-/instrument, a
front-end of CBMC), features transformations that can be applied to
the source program, such as k-induction for loops, constant
propagation and inlining. While our tool\-chain supports using those
CBMC features, none of them have proven effective in reducing the
complexity or analysis time in our WCET framework. In particular, the
loop acceleration included there takes longer than our entire WCET
analysis and does not improve the analysis time thereafter, and ``full
slicing'' even produces unsound results. Therefore, our source
transformations introduced above are justified, because they have a
proven impact on WCET analysis.

\subsection{Adding a Driver Function\label{sec:adding-driv-funct}}
To run a model checker on the instrumented (sliced, accelerated, abstracted) source code, we add a \emph{driver} function to the source that implements the following operations in sequence:
\begin{enumerate}
\item Initialize counter variable \texttt{\_time} to zero.
\item Initialize all input variables of the program to nondeterministic values according to their data type. This includes handling of persistent variables as described in Section~\ref{sec:handl-pers-vari}.
\item Call function $f$ for which WCET shall be estimated.
\item Encode assertion properties to query for WCET (details in Section~\ref{sec:wcet}).
\end{enumerate}

The driver function is handed over as entry point to the model checker.

\subsection{Determining WCET}\label{sec:wcet}
At this point, a model checker such as CBMC can verify whether the counter variable
\texttt{\_time} always carries a value less than $X$ after the program
terminates. In this section we explain how to choose candidates for $X$, such that we eventually approach the WCET.

A WCET candidate $X$ is encoded as \texttt{assert(\_time <= X)}, and
subsequently passed to the model checker. Unless the model checker
runs into a timeout or out of memory, only two outcomes are possible:
\begin{enumerate}
\item Successfully verified, i.e., \texttt{\_time} can never exceed $X$. Therefore, $X$ is a valid upper bound for the WCET.
\item Verification failed, i.e., \texttt{\_time} may exceed $X$
  in some executions. Therefore, $X$ is a lower bound for WCET. If a
  counterexample was generated, then it may contain a value $Y>X$,
  which then is a tighter lower bound for the WCET.
\end{enumerate}

Our strategy is to use both outcomes to narrow down on the WCET value
from both sides: Initially, we start with lower bound $e_\text{lower}$
as zero, and upper bound $e_\text{upper}$ as a very large value
(\emph{intmax}). We now place a number of assertions\footnote{We have empirically chosen $N_\text{assert}$=10; placing either more or less assertions usually take longer, because either the computational effort is growing, or more iterations are required.} 
in the
program, where each is querying one WCET candidate $X$. In
particular, the candidates are equidistantly spaced
between $e_\text{lower}$ and $e_\text{upper}$; except for the first step, 
where we use a logarithmic spacing to initially find the correct order
of magnitude. Subsequently, we invoke
the model checker to verify the assertions -- in the case of CBMC, all at
once. For each assertion we obtain a result. We set $e_\text{lower}$
as the largest $X$ where the assertion was failing (or, when a
counterexample with $Y>X$ was generated, to $Y$), and $e_\text{upper}$
as the smallest $X$ where the assertion was successfully verified. The
search is now repeated with the new bounds, and stopped when these
upper and lower bounds are close enough to each other, which can be
interpreted as a \emph{precision goal} for the WCET estimate.
The full algorithm is given in Algorithm~\ref{wcet-algo}.

\begin{algorithm}
  \DontPrintSemicolon
  \KwIn{instrumented C source code $C$, required precision $P$}
  \KwOut{WCET estimate $e_\text{upper}$,
s.t.~$e_\text{upper} - e_\text{lower} < P$.}
  \Begin{
    $N_\text{assert} \leftarrow 10$\tcp*{number of assert per call}
    $e_\text{lower} \leftarrow 0$\;
    $e_\text{upper} \leftarrow intmax$\;
    $p = e_\text{upper} - e_\text{lower}$\;
    \nl\While{$p>P$ \textbf{and not} timeout}{\label{InRes1}
      \If{$e_\text{upper} = intmax$}{
        \footnotesize candidates $\leftarrow$ logspace($e_\text{lower} .. e_\text{upper}, N_\text{assert}$)\;
      }
      \Else{
        \footnotesize candidates $\leftarrow$ linspace($e_\text{lower} .. e_\text{upper}, N_\text{assert}$)\;
      }
      \nl $C'\leftarrow$ insert \texttt{assert}s for candidates into $C$\;
      \nl results $\leftarrow$ model checker ($C'$)\;
      \For{$i = 1$ \KwTo $N_\text{assert}$}{
        \If{\emph{verified(results[i])}}{
          $e_\text{upper}\leftarrow \min(e_\text{upper}, \text{candidates[i]})$\;
        }
        \Else {
          \nl$B\leftarrow$ getCounterexample(results[i])\;
          $e_\text{lower}\leftarrow \max(e_\text{lower}, \text{candidates[i]},B)$\;
        }
      }
      \nl$p = e_\text{upper} - e_\text{lower}$\;
    }
  }
\caption{Iterative search for WCET bound\label{wcet-algo}}
\end{algorithm}

At any point in time the model checker could terminate due to either a
user-defined timeout or when it runs out of memory. In such cases the algorithm
returns the WCET estimate as at-this-point tightest bound
$e_\text{upper}$ that could be verified. In combination with the
precision goal $P$, this gives the user a fine control over how much
effort shall be spent on computing the WCET. For example, an
imprecise and fast estimate may be sufficient during early
development, whereas a precise analysis may be only of interest when the
WCET estimate approaches a certain time budget.

The maximum number of search iterations can be determined in advance;
in the worst case the number of search iterations $n$ is
\begin{equation}
  \quad n = \left\lceil
    \log_{N_\text{assert}}(\frac{e_\text{upper}-e_\text{lower}}{P})\right\rceil,
\end{equation}
where $e_\text{upper}=intmax$ and $e_\text{lower}=0$, if no a-priori
knowledge about the bounds of WCET is available. Usually the number of
iterations is lower, since the values found in the counterexamples
speed up the convergence (point 4 in Alg.~\ref{wcet-algo}).

\paragraph{Leveraging A-Priori Knowledge.} In this article we assume
that no information about the timing behavior of the program is
available.  If, however, the user has some knowledge on the WCET
bounds already, then these bounds can be tightened by the algorithm,
reducing the number of iterations.  If an upper bound is known, then
$e_\text{upper}$ can be initialized with that bound, and the algorithm
tightens it up to the required precision. Similarly, if a lower bound
is known from measuring the worst-case execution time on the target
(e.g., from a high watermark), then $e_\text{lower}$ can be set to that
measured value.

\paragraph{Implementation.}
The WCET search procedure was implemented as a Python script.
It further implements monitoring of memory and CPU time as given in
Table~\ref{table3}.  As model checker we used CBMC with various solver
backends. However, other model checkers, such as \emph{cpachecker}
could be used as alternatives with only little changes.

\subsubsection{Target-Specific Analysis} Since the analysis takes
place at source level, it is essential to include target-specific
information, such as word with, endianness, interrupts, I/O facilities
etc. If neglected, the behavior between the model in the analysis and
the real target may differ, leading to unsafe results. 
Most importantly, we provide target-specific preprocessor definitions
and the word widths with the corresponding flags that CBMC offers.
For more details on how to include target-specific information for the
model checker, we refer the reader to \cite{SAFECOMP15}, where the
specific pitfalls for CBMC have been explained. We further employ
checks during the WCET path reconstruction, which can identify missing or
incorrect target information.

\subsection{Determining BCET}
Our tool set can also be used to compute the Best-Case Execution Time
(BCET), with minor modifications. Such a lower bound of execution time may
be of interest for running a schedulability analysis of event-driven tasks, for
example, interrupts. Schedulability analysis then requires a \emph{minimum
  inter-arrival time} (MINT), i.e., the shortest possible time between
two consecutive releases, to bound the processing load that is
generated by the task.  If a software has no inherent mechanism to
limit the MINT of an event-driven task, then the BCET can be computed
and used in place of MINT.


\section{Reconstructing the WCET Trace\label{sec:replaying-wcet-trace}}
From a software developer's point of view, the WCET value in itself (and its inputs) are of limited use. Without understanding the exact path and the decision variables leading to the WCET, counteractive measures are limited to a time-consuming trial-and-error approach. Instead, the following information would be of high value for comprehension and proactive mitigation of the WCET case: 
\begin{enumerate}
\item The exact inputs leading to the WCET path,
\item a concrete walk-through of the worst\-/case path where all
  variables can be inspected, to identify the timing drivers, and
\item a timing profile of the worst\-/case path (how much time was spent where?).
\end{enumerate}
Especially a detailed walk-through of the worst-case path is very
important to understand why a certain path incurs high timing
cost. From our experience, it is not sufficient to know only the WCET
path. Oftentimes the specific values of decision variables are
important to understand why this path was taken, but such information
is not provided by any tool that we know of. The user is therefore
left with the mental puzzle of finding an explanation how the
presented path came to be, and at times there can be very subtle and
unintuitive reasons.  For example, in our \emph{prime} benchmark, we
found that an integer overflow in a loop condition caused the loop to
terminate only after a multiple of its explicit bound. In turn, this
overflow depends on the word width of the processor being used (more
details about this case are given in
Section~\ref{sec:wcet-path-reconstr}). In such a case a developer
would most likely struggle to explain why the loop did not obey to its
bounds, and thus not understand the WCET estimate.  In summary, the
WCET trace that we want to reconstruct shall be detailed enough to
inspect not only the control flow and its timing, but also all data
that is relevant for the specific path being taken. Towards that, we
want to leverage the final counterexample from Alg.~\ref{wcet-algo} to
provide a maximum-detail explanation of how the WCET can be reached.


\begin{figure*}
  \hspace{3mm}
    \begin{tabular}[t]{@{}lccr}
      \subfloat[Control flow graph\label{fig:wcetpath_cfg}]{%
        \includegraphics[width=.25\textwidth]{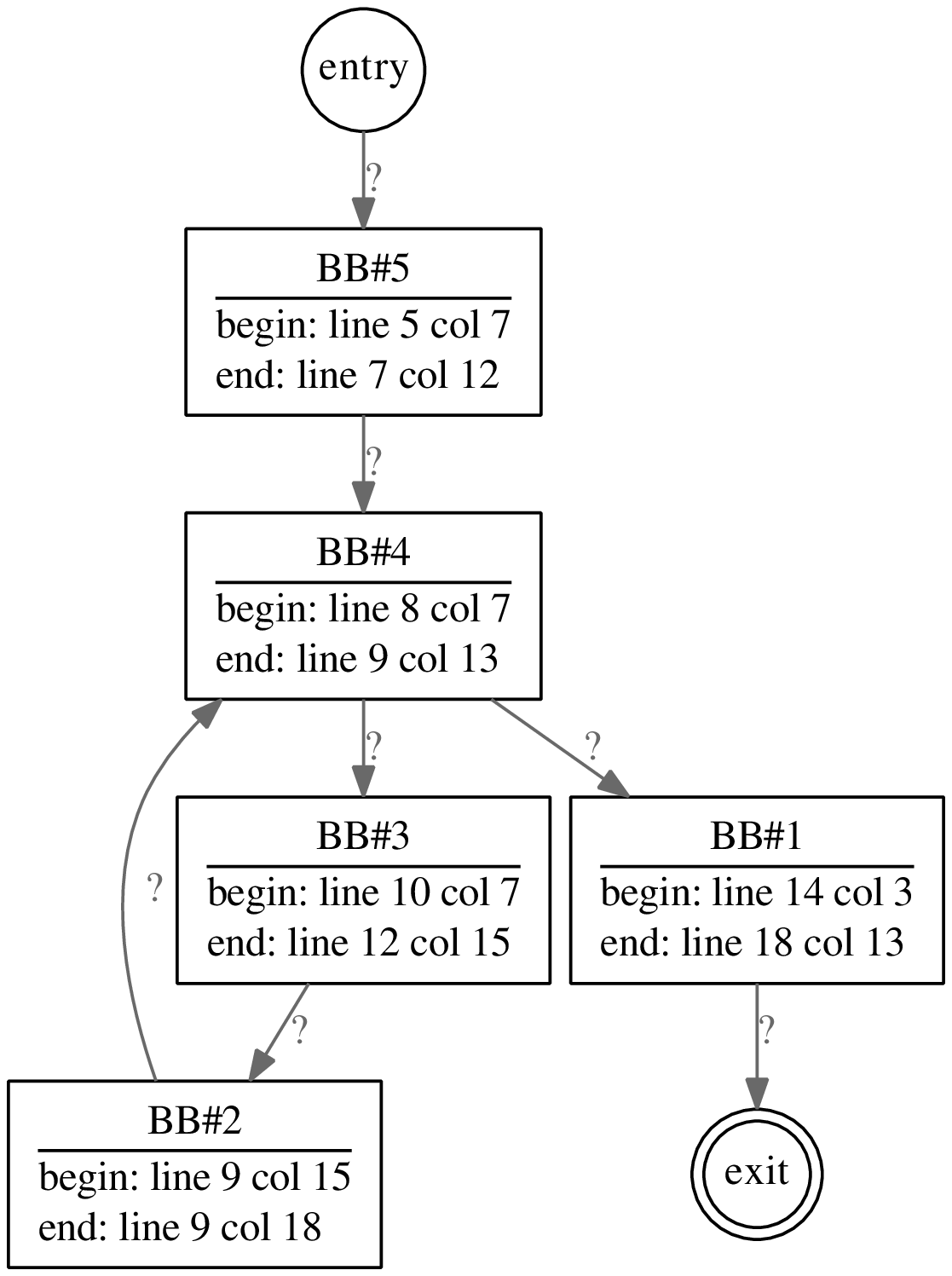}} \hspace*{1cm}&
      \subfloat[Typical output of model checker\label{fig:wcetpath_typical}]{%
        \includegraphics[width=.25\textwidth]{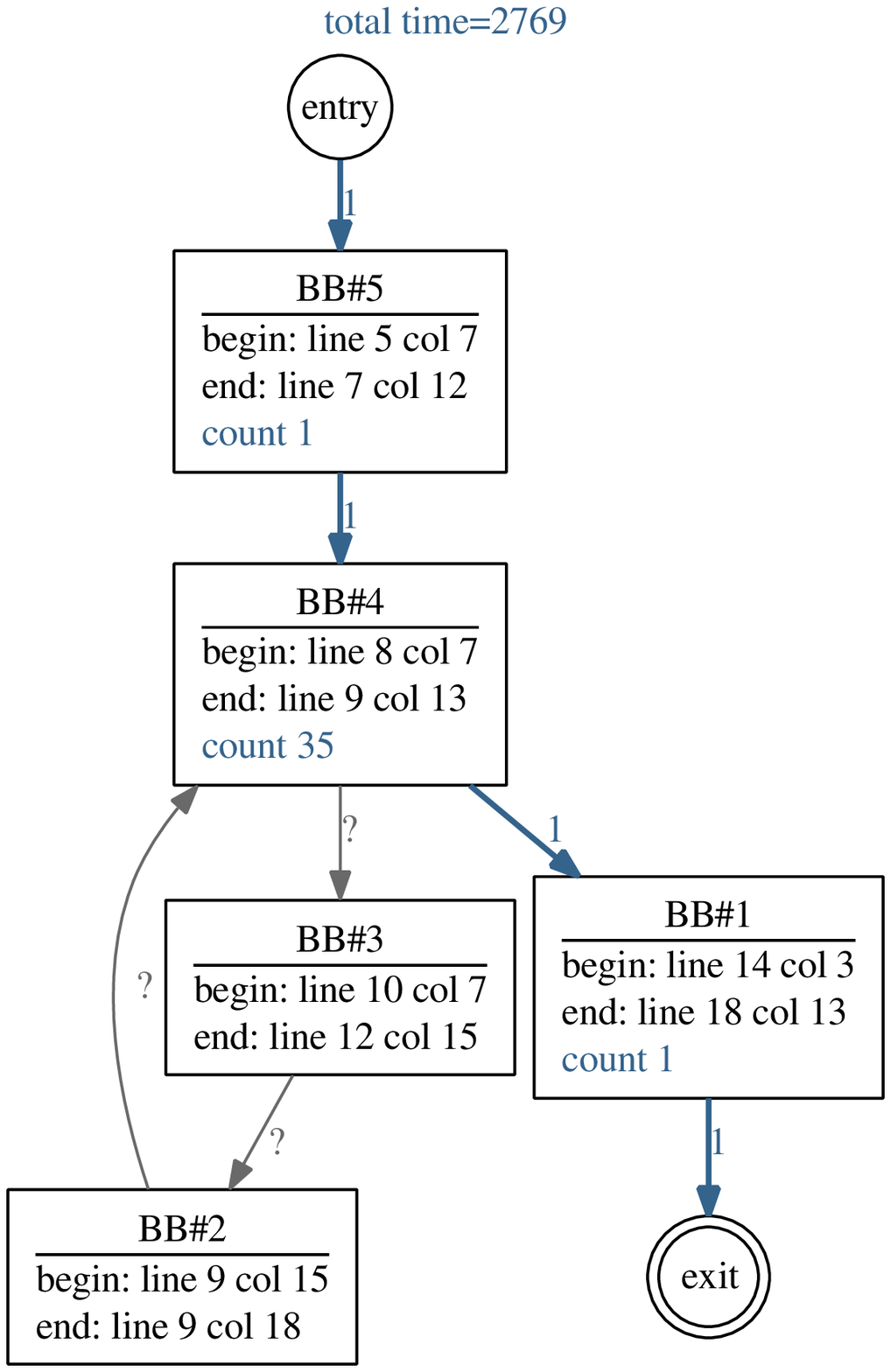}\hspace*{1cm}} &
      \subfloat[Fully reconstructed path\label{fig:wcetpath_full}]{%
        \includegraphics[width=.25\textwidth]{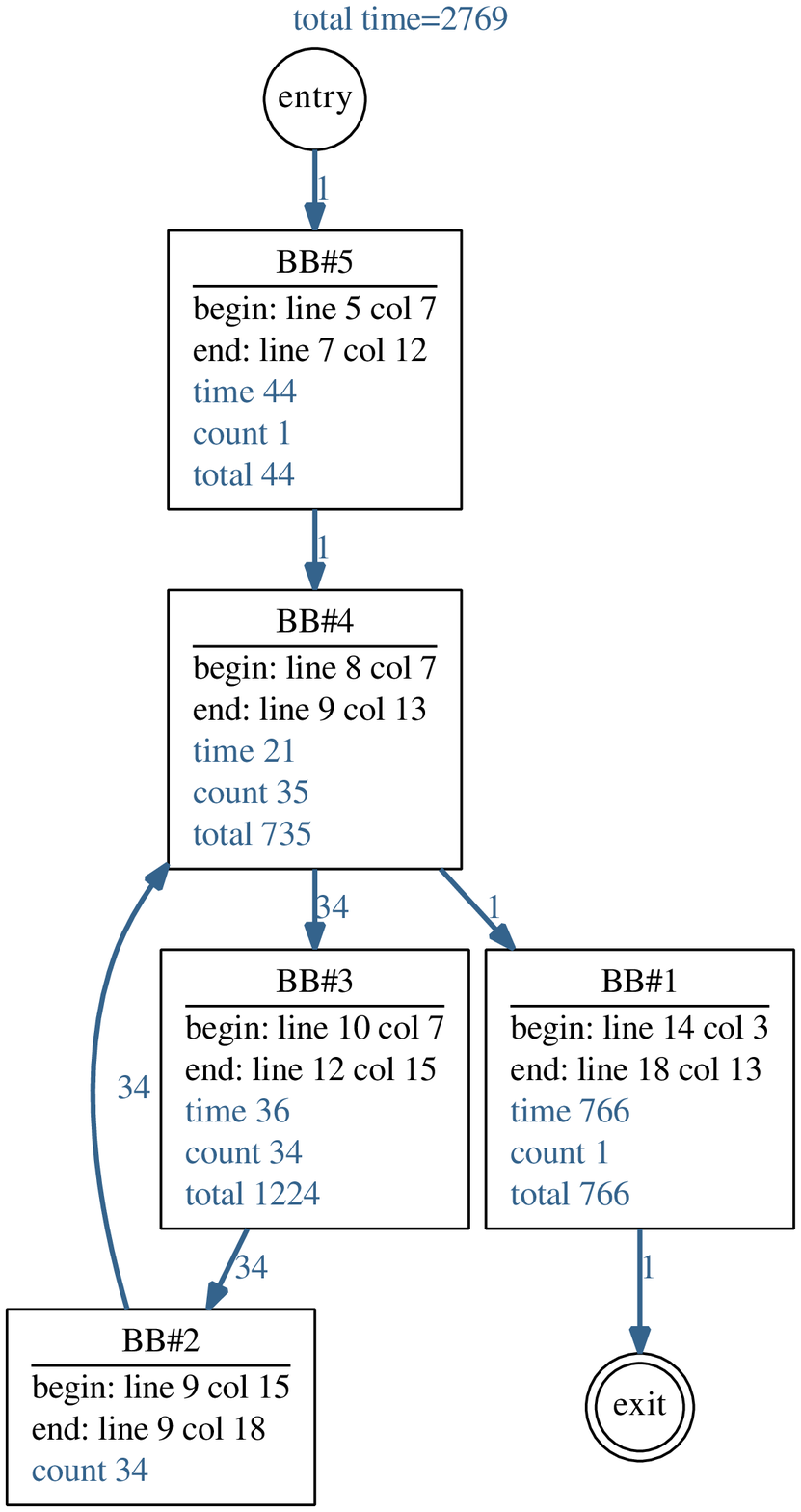}}%
    \end{tabular}
  \caption{Reconstruction problem of WCET path for example code from Figure~\ref{fig:orig}}
  \label{fig:wcetpath}
\end{figure*}

The challenge in reconstructing the WCET path from a
counterexample is illustrated in Figure~\ref{fig:wcetpath}, for the
same program from Fig.~\ref{fig:ins} introduced earlier. The goal is to
annotate 
the corresponding control flow graph in Fig.~\ref{fig:wcetpath_cfg} with execution
counts at both its nodes (basic blocks in source code) and its edges
(branches being taken). However, the counterexample
produced by the model checker only contains sparse information as
shown in Fig.~\ref{fig:wcetpath_typical}. Typically, it only provides
a subset of the visited code locations, variable assignments that are
relevant for branches being taken, and assignments to \texttt{\_time}
oftentimes occur only at the end of the counterexample, depending on
which solver backend was chosen. In fact, the counterexample is only
guaranteed to provide exactly one value for the variable
\texttt{\_time}, which is at the location where the WCET assertion is
failing.

It is clear that the counterexample provided by the model checker is
insufficient as an explanation for the WCET path, much less for values of arbitrary
variables. It could carry even less information than what is provided
by an ILP-based approach, where execution counts for all basic blocks
on the WCET path are available. Without such data, neither can we
compute a timing profile in the presence of function calls, nor is it
possible for the developer to understand or even walk through the WCET
path.

Therefore, towards reconstruction of the WCET path, we have to
interpolate the control flow in between the locations given in the
counterexample, and deduce variable valuations that are not available
(most importantly, variable \texttt{\_time}). There are two
fundamental approaches for reconstructing the path:
\begin{enumerate}
\item \textbf{By Analysis:} Use SMT, AI, or a similar technique to
  fill the ``location gaps'' of the counterexample, and to conclude
  about assignments of all (possibly sliced) variables. This is
  expected to be computationally complex and not precise.
\item \textbf{By Execution:} Execute the code with the worst\-/case
  inputs. An ``injection'' of critical values beyond inputs is
  required, for example to all persistent variables. This could be
  compiled into the application through look-up tables that contain
  the known assignments from the counterexample, or done dynamically
  during execution.
\end{enumerate}

It should be clear that an execution is preferable in terms of
precision and speed, however, there are some challenges in such an approach:
\begin{enumerate} 
\item \textbf{Using an Interpreter:} Whereas the most logical choice,
  only few good interpreters available for the C language. Most of
  them have limited ``stepping'' support (e.g., \emph{cling} uses JIT;
  which means stepping would inline/flattening functions), and in
  general they are slow. There would be no support for target-specific
  code, such as direct pointer dereferencing. Often only a subset of the
  language is implemented.
\item \textbf{Instrumented Execution:} Compile and run the
  application, preferably on the target for maximum fidelity, while
  capturing an execution trace which subsequently could be
  replayed. Capturing a complete trace including variable valuations
  could produce a huge amount of data and incur memory and timing
  overhead. If the program under analysis is supposed to run on a
  different target (cross-compilation), then it might not be feasible
  to capture the trace for reasons of memory limitations or missing
  interfaces.
\end{enumerate}

We decided for an execution-based approach, since this is computationally less complex than analysis, and thus expected to be faster. The problem of insufficiently granular interpretation and trace capturing has been addressed as follows: Our path reconstruction is accomplished by means of executing the application in a debugger, whilst injecting variable valuations from the counterexample when necessary. By choosing a debugger as replay framework, replaying the WCET has the potential to become intuitive to most developers, and thus could be seamlessly integrated into the software development process.
Furthermore, debuggers are readily available for most targets, and some processor simulators even can be debugged like a real target (in our case, \emph{simulavr} allows this). And, finally, word widths, endianness etc, are bound to be correct, in contrast to any approach based on an interpreter. However, to use these advantages to full capacity, the replay process must be of low overhead, and able to be fully automated to not require any additional inputs compared to traditional debugging.

\subsection{Preparing Replay}
\begin{figure}
  \centering
  \includegraphics[width=\columnwidth]{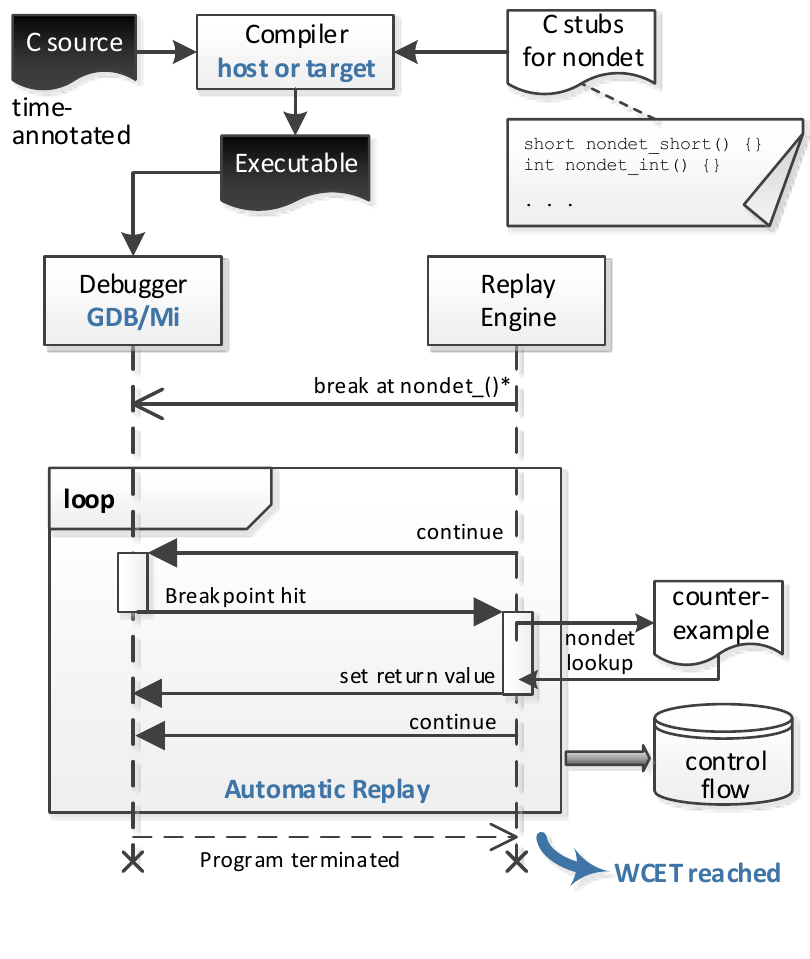}
  \caption{Automatic WCET replay using a debugger}
  \label{fig:replay}
\end{figure}

Our proposed replay process is illustrated in Figure~\ref{fig:replay}.
We start by compiling the instrumented (or sliced, accelerated,
abstracted) program with additional stubs (empty functions) for each
``nondet'' function (see section~\ref{Background:CBMC}). Then, we load
the program into a debugger.  Depending on which compiler was chosen,
this could either be the host's debugger, or the one for the target
(connecting to either an actual target or to a simulator). Since in
all cases the replay process is similar, for the rest of this paper we
do not distinguish anymore between a host-based replay, a simulator or
a real target.

Finally, it should be mentioned that the replay path is based on the time instrumentation 
in the sources, which is why the path always exists
unless the model checker is handed wrong target-specific settings
(e.g., word widths), or unless the analysis is unsound. 

\subsection{Injection of Nondeterminism}
To inject the critical variables as found by the model checker, we set breakpoints on the inserted nondet stubs and start the execution. As soon as the debugger reaches one of our breakpoints (additional user breakpoints for stepping may exist and are not interfering), we automatically inject the correct value as follows: First, we query the current call stack to identify the caller, i.e., the source location of the nondet call. Knowing the location of the designated nondet assignment, the value that leads to WCET is extracted from the counterexample, and subsequently forced as a return value in the debugger. As an effect, the call to the empty nondet stub returns the critical value suggested by the model checker. After that, the execution is resumed automatically. 

However, there exists a second source of non\-/determinism, besides the explicit ``nondet'' function calls. In the C language, every uninitialized variable that is not static nor at file scope, initially carries an unspecified value. Therefore, every such uninitialized local variable is considered by the model checker as non\-/deterministic input, as well. Since no explicit ``nondet'' function calls exist, the breakpoints inserted earlier do not allow us to inject values into those variables. As a solution, we first identify all uninitialized local variables from the parse tree of the C source (declarations without right-hand sides, mallocs, etc.), and then insert additional breakpoints for every such variable that is mentioned in the counterexample. Through this, injecting values into uninitialized variables is handled in the same way as the nondet function calls (not shown in the Fig.~\ref{fig:replay} for clarity).

With this technique, the injection of the critical variables from the counterexample is accomplished without any user interaction, and without embedding the assignments in the program itself (no memory overhead). Furthermore, this live injection allows for additional checks during the execution, such as matching the assignment with the actual call location, and ensuring that the execution path does not deviate from the analysis.
 
\subsection{Identification of WCET-Irrelevant Variables}
The valuations of some variables do not have an effect on the control
flow, and thus do not influence the timing\footnote{Note that this
  only holds true for cache-less processors.}. As explained before,
such variables are identified and sliced away during the analysis
phase. In particular, both our pre-processing (slicing, acceleration,
abstraction), as well as the model checker itself remove such
variables. Consequently, the counterexample does not include
valuations for variables that have been found irrelevant for the WCET.

As an effect, any location having a non\-/deterministic assignment that
is visited during replay \emph{and} and does not have an equivalent
assignment in the counterexample, indicates that the respective
assignment is irrelevant for WCET. We highlight such irrelevant
statements to the developer, to help focus on the drivers for the
worst\-/case timing, and not get distracted by surrounding code.

\subsection{Collecting a Timing Profile}
When larger programs are being analyzed, it may quickly become
impractical to step through the whole path to find the main drivers of
the worst\-/case execution time. To help the developer identify
interesting segments that need to be stepped through in the debugger,
we also generate a timing profile of the WCET path, showing which
location was visited how often, and how much time was spent there.

Towards that, we capture the complete control flow on the fly during
the replay. Since the timing profile is especially useful for larger
programs, capturing the control flow during the debugger execution
must scale well with growing path length and thus cannot be realized
by a slow step-by-step execution in the debugger. Instead, we set a
hardware \emph{watchpoint} on our counter variable. That
is, every time this variable is modified, the debugger pauses
execution and notifies the replay engine of the new variable
content. Since hardware watchpoints are realized through exceptions,
the debugger can run without polling and interruptions between the
watchpoints, and therefore the control flow is captured with very
little additional delay.  Considering that the counter variable is
embedded at least once per source block, 
the sequence of all reached watchpoints (their valuation and
location), represents the actual control flow in the source code.
As a result, a timing profile similar to the outputs of the well-known
tools \emph{gprof} or \emph{callgrind} can be reconstructed and
intuitively used by the developer. Table~\ref{tab:profile} shows the
resulting flat WCET timing profile for the \emph{adpcm decode}
benchmark. Note that additionally to the shown per-function metrics,
the execution times are also available at the even finer granularity
of source blocks, which helps pinpointing the timing bottlenecks to
specific source blocks within the functions.

\begin{table}
  \begin{center}
  \caption{Timing profile obtained from WCET path of \emph{adpcm decode} benchmark\label{tab:profile}}
  \vspace{-.5em}
\resizebox{\columnwidth}{!}{%
\pgfplotstabletypeset[
col sep=semicolon,
row sep=\\,
every head row/.style={
  before row={\toprule}, 
  after row={\midrule} 
},
header = true,
columns/name/.style={
  string type,
  column type=l,
},
columns/total/.style={
  column name=\%total,
  column type = r,
  fixed, precision=1, fixed zerofill
},
columns/tcycles/.style={
  column name=cycles,
  column type = r
},
columns/self/.style={
  column name=\%self,
  column type = r
},
columns/scycles/.style={
  column name=cycles,
  column type = r
},
columns/calls/.style={
  column type = r
},
columns/scall/.style={
  column name=self/call,
  column type = r
},
columns/tcall/.style={
  column name=total/call,
  column type = r
},
every last row/.style={after row=\arrayrulecolor{black}\bottomrule}
]{
total; tcycles; self; scycles; calls; scall; tcall; name\\
100.0;69673;35.3;24593;1;24593;69673;decode\\
13.7;9522;13.7;9522;2;4761;4761;upzero\\
13.2;9168;13.2;9168;2;4584;4584;uppol2\\
12.9;8984;12.9;8984;2;4492;4492;filtez\\
9.3;6472;9.3;6472;2;3236;3236;uppol1\\
5.5;3854;5.5;3854;2;1927;1927;filtep\\
5.1;3576;5.1;3576;2;1788;1788;scalel\\
2.5;1758;2.5;1758;1;1758;1758;logscl\\
2.5;1746;2.5;1746;1;1746;1746;logsch\\
}
}
  \end{center}
\end{table}



\section{Experiments}
\label{expts}

We applied \tick\ to the M\"alardalen WCET Benchmark Suite
\cite{Gustafsson:WCET2010:Benchmarks} to evaluate the performance and the tightness of 
WCET estimates computed with \tick.  As a target, we used the Atmel ATmega 128~\cite{AVRdatabook}, for which WCET analyzers (Bound-T~\cite{holsti2002status}) and simulators (simulavr)
are freely available and can be used as a baseline for evaluating \tick. This target satisfies our requirements for WCET\-/amenable processors, since there is practically no timing dependence on the operands.

The complete set of experimental data (instrumented, sliced,
accelerated and abstracted sources, as well as complete traces of the WCET search) is
available at \benchmarksurl.

\subsection{Setup and Reference Data}
\paragraph{Selected Benchmarks.} 
We selected a representative subset consisting of 17 M\"alardalen WCET benchmarks, such that all program properties, e.g., multi-path flows,
nested loops, arrays and bit operations were covered, except for
recursion and unstructured code (they cannot be handled by the basic
block extractor, yet), and floating point variables (cannot be handled
by the timing instrumentor, yet). However, these missing properties in
principle can be addressed; this is not a limitation of our approach.
 
\paragraph{Host Platform.} 
We conducted our experiments on a 64bit machine with a 2.7GHz Intel Xeon E5-2680 processor and 16GB RAM, using CBMC 5.6 as model checker. As CBMC's backend we have used a portfolio of solvers, consisting of minisat (built-in), mathsat (v5.3.10/\-GMP), cvc4 (v.1.5pre/\-GLPK-cut), z3 (v.4.5.1) and yices (v.2.5.1/\-GMP). We stopped the analysis as soon as the first solver provided a WCET estimate. Solvers finishing within the same second were considered equally fast. All programs have been analyzed sequentially, to minimize interference between the analyses and with third-party background processes. The computational effort (CPU time, peak memory usage) are derived from the Linux system call \texttt{wait3}, and thus expected to be accurate.

\paragraph{Bounding the Control Flow.} In most cases our approach could bound the control flow automatically, and no manual input was required. However, in two benchmarks, namely \emph{bs} and \emph{insertsort}, CBMC could not find the bounds automatically and we were not able to accelerate or abstract either, and thus bounds had to provided. This occasional need for manual bounds is discussed in detail in Section~\ref{sec:usability-safety}.


In contrast, we frequently had to provide loop bounds for the ILP-based WCET analyzer. Since in an ILP approach the bounds cannot be verified, they have to be specified correctly and tightly in the first place. Towards this, whenever the ILP-based estimation required manual loop annotations, we have taken the deduced bounds from our approach and handed them to the ILP-based analyzer. Consequently, both techniques had similar preconditions for their WCET estimation.

\paragraph{Simulation Baseline.}
All benchmarks were also simulated with the cycle-accurate ISS \emph{simulavr}, with the goal of having sanity checks for the WCET estimates, and also to get an impression on their tightness. Whenever possible, the simulation was performed with those inputs triggering the WCET. In other cases, we used random simulations in an attempt to trigger the WCET, but naturally we cannot quantify how close to the actual WCET we have come (e.g., in nsichneu). Hence, the simulation results can only serve as a lower bound for the actual WCET (i.e., no estimate must be less) and as an upper bound for the tightness (i.e., the overestimation of the actual WCET is at most the difference to the simulation value).

\begin{table}[t]
  \begin{center}
    \caption{Tightest WCET estimates per method and benchmark (timeout 1 hour)\vspace{-.75em}}
    \label{table1}
\resizebox{\columnwidth}{!}{%
  \begin{tabular}[htb]{llrrrrr}
    \toprule
                   &           Simulation & \multicolumn{2}{c}{ILP-based} & \multicolumn{3}{c}{Model Checking}\\
    \cmidrule(rl){2-2}\cmidrule(rl){3-4} \cmidrule(rl){5-7}
    benchmark           & observed   & WCET           & $\Delta\%$        & stage & WCET & $\Delta\%$ \\
    \midrule
    \arrayrulecolor{gray!50}
adpcm-decode & 48,168 & 71,575 & +48.6 & \instrumented & 69,673 & +44.6\\
 &  &  &  & \sliced & 69,673 & +44.6 \\
 &  &  &  & \accelerated & 69,673 & +44.6\\
 &  &  &  & \abstracted & 69,673 & +44.6\\
\specialrule{.2pt}{1pt}{2pt}adpcm-encode & 72,638 & 113,154 & +55.8 & \instrumented & 110,901 & +52.7\\
 &  &  &  & \sliced & 110,901 & +52.7\\
 &  &  &  & \accelerated & 110,901 & +52.7\\
 &  &  &  & \abstracted & 110,901 & +52.7\\
\specialrule{.2pt}{1pt}{2pt}bs & 401 & 496 & +23.7 & \instrumented & 410 & $\approx$0\\
\specialrule{.2pt}{1pt}{2pt}bsort100 & 788,766 & 1,553,661 & +97.0 & \instrumented & timeout & --\\
 &  &  &  & \abstracted & 797,598 & +1.1\\
\specialrule{.2pt}{1pt}{2pt}cnt & 8,502 & 8,564 & $\approx$ 0 & \instrumented & 8,564 & $\approx$0\\
 &  &  &  & \sliced & 8,564 & $\approx$0\\
 &  &  &  & \abstracted & 8,564 & $\approx$0\\
\specialrule{.2pt}{1pt}{2pt}crc & 129,470 & 143,137 & +10.6 & \instrumented & 130,114 & $\approx$0\\
 &  &  &  & \sliced & 130,114 & $\approx$0\\
 &  &  &  & \accelerated & 130,114 & $\approx$0\\
 &  &  &  & \abstracted & 143,426 & +10.8\\
\specialrule{.2pt}{1pt}{2pt}fdct & 17,500 & 17,504 & $\approx$ 0 & \instrumented & 17,504 & $\approx$0\\
 &  &  &  & \sliced & 17,504 & $\approx$0\\
 &  &  &  & \accelerated & 17,504 & $\approx$0\\
\specialrule{.2pt}{1pt}{2pt}fibcall & 1,777 & 1,781 & $\approx$ 0 & \instrumented & 1,780 & $\approx$0\\
 &  &  &  & \sliced & 1,780 & $\approx$0\\
 &  &  &  & \accelerated & 1,780 & $\approx$0\\
\specialrule{.2pt}{1pt}{2pt}fir & 5,204,167 & 5,690,524 & +9.3 & \instrumented & timeout & --\\
 &  &  &  & \sliced & timeout & --\\
 &  &  &  & \accelerated & 5,476,023 & +5.2\\
\specialrule{.2pt}{1pt}{2pt}insertsort & 5,472 & 5,476 & $\approx$ 0 & \instrumented & 5,476 & $\approx$0\\
\specialrule{.2pt}{1pt}{2pt}jfdctint & 14,050 & 14,054 & $\approx$ 0 & \instrumented & 14,054 & $\approx$0\\
 &  &  &  & \sliced & 14,054 & $\approx$0\\
 &  &  &  & \accelerated & 14,054 & $\approx$0\\
\specialrule{.2pt}{1pt}{2pt}matmult & 1,010,390 & 1,010,394 & $\approx$ 0 & \instrumented & 1,010,394 & $\approx$0\\
 &  &  &  & \sliced & 1,010,394 & $\approx$0\\
 &  &  &  & \accelerated & 1,010,394 & $\approx$0\\
\specialrule{.2pt}{1pt}{2pt}ndes & 459,967 & 470,499 & $\approx$ 0 & \instrumented & timeout & --\\
 &  &  &  & \sliced & timeout & --\\
 &  &  &  & \abstracted & 465,459 & $\approx$0\\
\specialrule{.2pt}{1pt}{2pt}ns & 56,409 & 56,450 & $\approx$ 0 & \instrumented & 56,413 & $\approx$0\\
 &  &  &  & \abstracted & 56,450 & $\approx$0\\
\specialrule{.2pt}{1pt}{2pt}nsichneu & 33,199 & timeout & -- & \instrumented& timeout & --\\
 &  &  &  & \abstracted & 75,369 & +127.0\\
\specialrule{.2pt}{1pt}{2pt}prime & 27,702,943 & 30,343,092 & +9.5 & \instrumented & timeout & --\\
 &  &  &  & \sliced & timeout & --\\
 &  &  &  & \abstracted & 30,146,785 & +8.8\\
\specialrule{.2pt}{1pt}{2pt}ud & 35,753 & 93,487 & +161.5 & \instrumented & 38,992 & +9.1\\
 &  &  &  & \sliced & 38,992 & +9.1\\
 &  &  &  & \accelerated & 38,992 & +9.1\\
 \arrayrulecolor{black}\bottomrule
 \multicolumn{7}{c}{\tablelegend, $\Delta$~upper bound for tightness}
  \end{tabular}
}

  \end{center}
\end{table}
 
\begin{table}[t]
  \begin{center}
    \caption{Complexity and computational effort of WCET search for precision of 10,000 clock cycles and timeout of 1min}
    \label{table3}
\providetoggle{oddbench}%
\toggletrue{oddbench}%
\pgfplotstableread[col sep=semicolon]{results_cbmc_portfolio.csv}\loadedtable 
\pgfplotstablecreatecol[%
create col/assign/.code={%
  \getthisrow{iterations(precision=10000)}\iter%
  \edef\entry{\iter}
  \pgfkeyslet{/pgfplots/table/create col/next content}\iter%
}
]{iterations}\loadedtable
\pgfplotstablecreatecol[%
create col/assign/.code={%
  \getthisrow{peak memory winner}\pmemw%
  \getthisrow{peak memory portfolio}\pmemp%
  \expandafter\ifstrequal\expandafter{\pmemw}{}{%
    \edef\entry{\pmemp}%
    \pgfkeyslet{/pgfplots/table/create col/next content}\entry%
  }{%
    \edef\entry{\pmemw}
    \pgfkeyslet{/pgfplots/table/create col/next content}\entry
  }
}
]{peakmem}\loadedtable
\resizebox{\columnwidth}{!}{%
\hspace{-4.6cm}
\pgfplotstabletypeset[
every head row/.style={
  before row={\toprule}, 
  after row={\midrule} 
},
columns={name,version,winning solvers,iterations,best runtime,peakmem,program steps avg},
header = true,
string replace*={None}{},
string replace*={finder-i}{\instrumented},
string replace*={finder-s}{\sliced},
string replace*={finder-acc-auto-2}{\accelerated},
string replace*={finder-abs-auto-2}{\abstracted},
columns/winning solvers/.style={
  column type={l},
  string type,
  column name=fastest solver(s),
  string replace*={minisat}{A},
  string replace*={mathsat}{B},
  string replace*={yices}{C},
  string replace*={z3}{D},
  string replace*={cvc4}{E},
  empty cells with={timeout},
},
columns/best runtime/.style={
  column name = {time [s]},
  column type = r,
  fixed,precision=1,fixed zerofill,
  empty cells with={timeout}
},
columns/BoundT_time/.style={
  column name = {ILP time [s]},  
  column type = {r},
  string type,
  empty cells with={\ensuremath{-}}
},
columns/version/.style={
  string type,
  column type=c,
  string replace*={/wcet.log}{},
  string replace*={/wcet-cbmc}{},
  column name=stage
},
columns/name/.style={ 
  column type={l},
  string replace*={/wcet.log}{},
  string replace*={/wcet-cbmc}{},
  column name=benchmark,
  string type,
  empty cells with={},
  postproc cell content/.prefix code={
    \ifx\\##1\\
    \else      
      \iftoggle{oddbench}{\togglefalse{oddbench}}{\toggletrue{oddbench}}
        \ifnum\pgfplotstablerow=0
        \else
          \pgfkeyssetvalue{/pgfplots/table/@cell content}{          
            \arrayrulecolor{gray!50}\specialrule{.2pt}{1pt}{2pt} ##1
          }        \fi
    \fi
  }
},
columns/iterations/.style={
  column name = iter.,
  column type=r,
  empty cells with={1}
},
columns/BoundT_peakmem/.style={
  column name = ILP mem [MB],
  divide by=1048576, 
  column type=r,
  int trunc,
  empty cells with={TODO},
},
columns/peakmem/.style={
  column name = mem [MB],
  divide by=1048576, 
  column type=r,
  int trunc,
  empty cells with={TODO},
},
columns/program steps avg/.style={
  column name = prog.size,
  column type=r,
  empty cells with={\ensuremath{-}},
},
columns/iterations/.style={
  column name = iter.,
  column type=r,
  empty cells with={--}
},
every last row/.style={after row=\arrayrulecolor{black}\bottomrule\multicolumn{7}{c}{Step: \tablelegend}\\\multicolumn{7}{c}{Solvers: A=minisat, B=mathsat, C=yices, D=z3, E=cvc4}}, 
]\loadedtable
}
  \end{center}
\end{table}

\subsection{Results} 

Tables~\ref{table1} and \ref{table3} summarize our experiments. We evaluated our technique of WCET estimation for each source processing stage (Section~\ref{sec:step-1:-backtr}) of the selected benchmarks, that is:
\begin{samepage}
\begin{enumerate}
\item instrumented with execution times (\instrumented),
\item sliced w.r.t. timing (\sliced), 
\item loops accelerated (\accelerated) and
\item abstracted (\abstracted). 
\end{enumerate}
\end{samepage}

In Table~\ref{table1} we compare the tightness of our WCET estimate with that of Bound-T, an ILP-based WCET analyzer. We computed the tightest possible WCET, i.e., precision $P=1$, while allowing for a (relatively long) timeout of one hour, to show how our source transformations influence the tightness. Cases denoted with \emph{timeout} are those where no solution was found within that time budget. 
 Finally, the columns denoted as $\Delta$ represent the difference between the respective WCET estimates and simulation value, i.e., they give an upper bound of the tightness for both techniques. 

Table~\ref{table3} summarizes the computational effort of the estimation process for a practical time budget of one minute and a precision of 10,000 clock cycles.
The table also quantifies the speedup we have achieved with our source transformation techniques. Again we denote \emph{timeout} in cases where the WCET could either not be bounded within the time budget, or not up to the required precision.
The column \emph{prog.size} shows the number of program steps found by CBMC. 
Cases where the program size is not given (viz., in \emph{fir} and \emph{prime}) indicate a state-space explosion. That is, the model checker never finished constructing the state space before the timeout. 
The column \emph{iter} denotes the number of iterations of the search procedure (Algorithm~\ref{wcet-algo}) and
the column \emph{time} denotes the total time taken by the search procedure in seconds.
Cases with a valid program size \emph{and} a timeout (e.g., \emph{bsort100} \instrumented) indicate that the solver backend could not verify the given properties within 10 minutes. 

Finally, some benchmarks in Tables \ref{table1} and \ref{table3} are lacking a sliced, accelerated or abstracted version. This occurs when the respective source transformation technique did not result in any changes to the source code. For example, \emph{bsort100} remained unmodified post slicing and acceleration and thus does not have a dedicated sliced or accelerated version.

\subsection{WCET Path Reconstruction\label{sec:wcet-path-reconstr}}
We were able to reconstruct and replay the WCET path for all
benchmarks. The time taken for the debugger-based replay is in the
range of a few seconds in all cases. For a better usability, we
enabled our replay engine to output the trace as a CBMC-like
counterexample, but additionally augmented with all assignments to
\texttt{\_time}. Recall
that the complete information about variable \texttt{\_time}
implicitly carries the control flow, because each block in the source
code increments this variable. As an effect, the graphical user
interface can load this augmented counterexample, and map back the
counterexample to the control flow graph, as well as compute a timing
profile at a granularity of block- or function-level.

\begin{figure*}
   \centering
   \includegraphics[width=\textwidth]{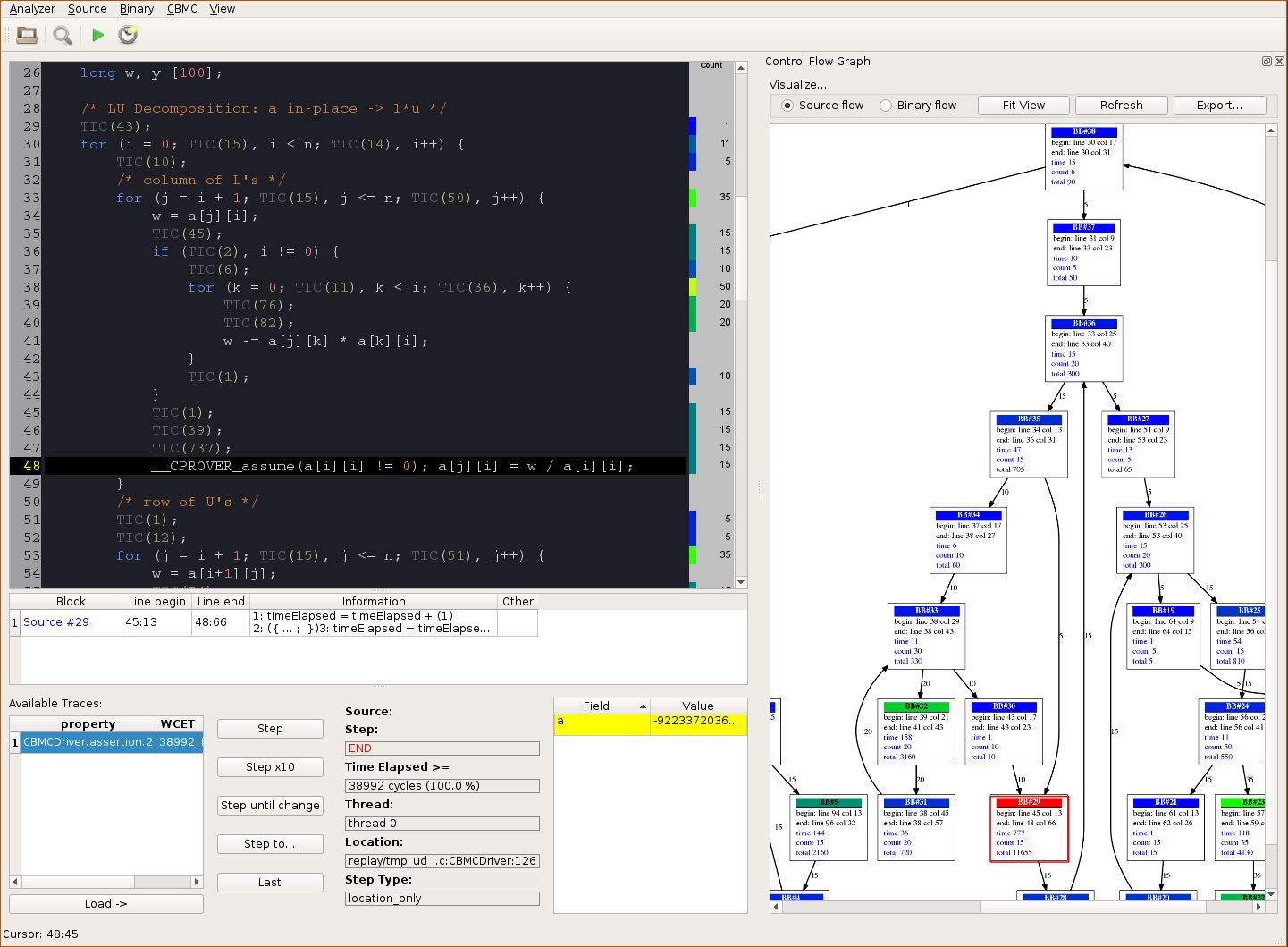}
   \caption{WCET path reconstruction for benchmark \emph{ud} with identified timing hotspot}
   \label{fig:gui}
\end{figure*}

\paragraph{Identifying a Timing Hotspot.}
As an example, we show the resulting annotated 
control flow graph for the \emph{ud} benchmark in
Figure~\ref{fig:gui}. There, we have applied a heatmap over timing,
where red marks the timing hotspots. At first glance, it is apparent
that there exists one source block forming a timing bottleneck, which
consumes about one third of the overall execution time.

The \emph{ud} program performs a simple form of LU decomposition of a
matrix $A$, and subsequently solves an equation system $Ax=b$. The
timing bottleneck in this program occurs in the LU decomposition, but
interestingly not at the innermost or most frequently executed loop,
but at a location where a division occurs. With this, the
reconstruction of the WCET path made it easy to spot that for the
chosen target, a long division incurs a high execution cost (see also
the inline annotations \texttt{TIC}), and that this is the single
driver for the WCET of this program.

\paragraph{Discovering a Bug.} The replay of the WCET path can also
reveal defects in the program. Such a defect has been found in the
\emph{prime} benchmark on a (simulated) 32-bit target, where initially
the WCET estimate just seemed suspiciously high. After reconstructing
the WCET path, an unexpected implausibility showed up. The path was
indicating that the following
loop had been executed $349,865$ times:\\
\hspace*{2em}\begin{minipage}{.9\linewidth}
\begin{lstlisting}[basicstyle=\small\ttfamily]
for (uint i=3; i*i <= n; i+=2) { 
  if (divides (i, n)) return 0;
  ... 
}
\end{lstlisting}    
\end{minipage}\\
where \texttt{n} is a formal parameter of the surrounding function
that shall be checked for being a prime number.  At first glance it
seems like the loop can only execute $65,536$ times: the maximum
32-bit unsigned integer \texttt{n} that the user can provide is
$4,294,967,295$, therefore the loop must execute at most
$\left\lceil\sqrt{4,294,967,295}~\right\rceil=65,536$ times to
evaluate whether that number is prime. Thus, we replayed the WCET path
in our debugger, setting a watchpoint on \texttt{i}. It turned out
that the expression \texttt{i*i} will overflow for some specific
values of $\text{\texttt{n}}>4,294,836,225$, since this would require
at least \texttt{i}$\geq 65,536$, which, when squared, lies beyond the
maximum representable unsigned integer and thus
overflows. Specifically, this happens for only those numbers which are
also not divisible by anything less than $65,536$, hard to replicate
if only the path and no values would be available.  As an effect, the
loop keeps running until finally the 32-bit modulo of \texttt{i*i} is
larger than \texttt{n} before the loop terminates (which luckily is
always the case). Clearly, this is a defect in the context of this
program, resulting in a WCET which was orders of magnitude higher than
it should have been\footnote{Note that this code works correctly for
  16-bit targets, as $\forall n \in \text{\texttt{uint16}}, \exists i
  \leq 255, \text{~s.t. \texttt{divides(i,n)}}$.}. Thanks to the ability of interactively replaying
the WCET path and inspecting variable values, such defects can be
easily spotted.

\paragraph{Top Ten Longest Paths.} In principle our approach could
also provide the ``top 10'' longest paths in the program. This could
be done by repeating the overall WCET estimation, while excluding
already known paths from the solution, and decreasing the WCET
proposal to the Model Checker if no counterexample can be
found. However, then this would become an explicit enumeration of
paths, typically exponentially in the size of the
program~\cite{Li1997}, and would still leave us with the question of
how the program's overall probability distribution looks like (unless
we repeat our estimation until every possible path length has been
found). We did not investigate this further, as this clearly would not
scale well with program size.



\section{Discussion}
Using our approach, the WCET could be estimated for all benchmarks, within a few seconds,
including a timing profile and path replay of the WCET case. Through that, not only did we 
provide an estimate without minimal user inputs, but also we generate useful
insights into the WCET path, enabling a true ``time debugging''.
As we elaborate on the following pages, source transformation techniques had a noticeable impact on scalability of Model Checking, and thus paved the way to use Model Checking for WCET analysis. As a result, the estimates are comparable to those of an ILP-based technique, but with less effort and more
feedback for the user.

\subsection{Tightness}
The tightness of \tick's WCET estimate can be expected to be almost exact
when Model Checking is allowed to explore all paths (i.e., no
abstraction applied), and becoming less tight when abstractions are
applied, comparable to an ILP-based solver where only loops are
bounded with no further flow analysis. In fact, Table~\ref{table1}
shows that our WCET estimates often are even tighter than an ILP-based
approach.

For the \emph{adpcm} benchmarks, there is still a lot of room for
improvement. The computed WCET, even on the original version, is
likely a large overestimation, as suggested by the simulation. The
reasons for this are discussed in the following.

\subsubsection{Intrinsic Sources of Overapproximation\label{sec:sourc-over}} 
During the back translation of timing information from machine instructions to
source code, overapproximations have been used as described earlier.  We
argue that these over-approximations are common even when using the
existing ILP-based techniques. During back translation, wherever there
is a difference between the source blocks and basic blocks in the machine instructions, we
over-approximate that part machine instructions into one block. 
However, these over-approximations are usually small, since these differences are 
often formed by few and small basic blocks, amounting to only a few clock cycles per iteration. 

Without any abstractions, \tick\ will exhaustively explore all
feasible paths in the (instrumented) source code, and therefore not
make additional overapproximations. Thus, when identifying the longest path,
\tick\ will never be worse than an ILP-based path search. In fact, the
ILP solver could only perform better, if the control flow could be
bounded \emph{exactly} and then encoded in the ILP formulation.

A typical case is that of \emph{crc}. When no abstractions were
applied, our estimate is around 10\% tighter than that of Bound-T
($130,114$ vs. $143,137$), which we tracked down to infeasible paths being considered by Bound-T: In \emph{crc},
there exists a loop containing an if-else statement, whereas the
if-part takes more time than the else-part. Bound-T, not analyzing the
data flow and dependency between these two parts, always assumes the
longer if-branch is taken. However, this is only feasible 70\% of the
time, whereas the other 30\% the shorter else-branch takes
place. Without additional user annotations or a flow analysis, ILP
cannot discover this dependency and thus must overestimate the WCET.

Consequently, \tick\ performs better than an ILP-based approach when
abstractions are left aside, if the program complexity had been
reduced enough by slicing and acceleration. For the remaining cases,
where abstraction was necessary to make Model Checking scale, we now
must discuss the overestimation caused thereby.

\subsubsection{Overestimation due to Abstraction}
Abstraction overapproximates the control and data flows, trading
analysis time for tightness. When applying loop abstraction, the
abstracted code forces the model checker to pick times along the
branch with the longest times, cutting out shorter branches. Thus, we
compute the WCET assuming that every branch inside the loop will
always take the worst local choice, which may be
infeasible. Naturally, this leads to an overestimation of
WCET. However, ILP-based analyzers usually over-approximate with
similar strategies when bounding the control flow. The result is a
WCET estimate comparable to that of an ILP-based analyzer.

Again, consider \emph{crc} as a typical case. When no abstractions
were applied, our estimate was around 10\% tighter than that of
Bound-T. When applying abstractions, the estimate became very close to
the estimate of Bound-T, whereas the complexity was cut down to
approximately 25\%.

Surprisingly, in some benchmarks, namely \emph{adpcm-decode} and \emph{encode},
\emph{cnt}, \emph{fdct}, \emph{jfdctint} and \emph{ud},
the loop abstraction did not lead to a higher WCET estimate. This is
because in all the loops with branches in these programs (a)~either
there is an input to the program that forces the branch with the
highest time to be taken in all the iterations of the loop, or
(b)~there is a break or return statement that gets taken in the last
iteration of the loop. These cases match the exact pessimistic loop
abstraction. In short, these loops do exhibit the pessimistic worst
case timing behavior.

An extreme overestimation due to abstraction seems likely (reminder:
the simulation is not guaranteed to contain the worst-case) for
\emph{nsichneu}, where the estimate is 127\% higher than the observed
WCET. This benchmark has only one loop with two iterations, but its
simulated WCET is far away from the observed WCET. Upon inspection, we
found that this loop has 256 if-statements in a single sequence, many
of which do not get executed in every iteration. However, our loop
abstraction pessimistically assumes that all the 256 if-statements do
get executed in each iteration, which explains the overestimation in
this case. Note that this benchmark could not be solved with Bound-T,
at all.

Consequently, \tick~performs close to and often slightly better than an
ILP-based approach when abstractions are used. However, this result
depends on the way control flows are bounded before the ILP solver is
called, and thus it might not hold true when AI and ILP are combined.

\subsection{Reduction of Computational Effort} 
Our claim was, that the scalability issues of Model Checking could be
mitigated with appropriate source preprocessing techniques, which we
expected to be particularly effective at source code level. The
experimental data clearly confirms that claim. In all cases, the
analysis time -- usually in the range of minutes up to unsolvable size
for the original version -- could be reduced to an acceptable time of
at most a few seconds, which makes \tick\ an interesting alternative
to existing WCET estimation
techniques.

However, taking the analysis time (computational effort) as measure of the computational
complexity can be misleading. 
The time to solve the WCET problem in our approach consists of two
fundamental parts:
\begin{inparaenum}
\item building the SAT/SMT model and
\item solving the model.
\end{inparaenum}
Whereas the time to build the model is often proportional to the
size of the program (number of steps as found by CBMC), the time for solving
the model cannot be predicted trivially by looking at program
metrics. In particular, we have found no correlation between any of 
program size, cyclomatic complexity, number of
statements, number of loops and the analysis time. The reason for
this is that the analysis time of a SAT or SMT problem can vary
significantly between two problems of the same size or between two solvers,
due to the nature of modern solver algorithms~\cite{Demyanova2015}. For
instance, compare \emph{adpcm-encode}\instrumented~(program size
$4,911$ steps, timeout after 1 minute) with
\emph{ndes}\abstracted~(program size $5,727$ steps, solved in less
than one second).

Therefore, we used a portfolio of solvers to reduce the effect of
solver-specific strengths and weaknesses, and we consider the program size (last column in Table~\ref{table3}) as a prime indicator
for the complexity of the program. With this, we show in the following that the 
complexity of a program can be significantly reduced with our techniques.  Nevertheless, note
that the solving time also points towards our interpretation. In all
benchmarks we were able to greatly reduce the computational effort with
each source processing stage.  In particular, several benchmarks could not be
solved in a reasonable amount of time without our proposed processing, viz., \emph{adpcm-encode},
\emph{bsort100}, \emph{fir}, \emph{ndes}, \emph{nsichneu} and
\emph{prime}. The memory usage suggests, that a state-space explosion
prevents building the model. Their program size could be reduced to a
tractable size with acceleration and abstraction.  Moreover, the
benchmark \emph{nsichneu} could not even be processed with Bound-T,
since bounding of the control flow had failed because of too many
variables and flows (out of memory after running for several
hours). After applying our abstraction, the WCET of this benchmark
could be successfully estimated with Model Checking, within one second
of analysis time.

\begin{figure}
  \centering
  \pgfplotsset{compat=1.8}
\usepgfplotslibrary{statistics}
\begin{tikzpicture}
\begin{axis}
    [
    boxplot/draw direction=y,
    boxplot/box extend=0.5,
    ymin=-0.1,enlarge y limits={0.1,upper},
    ylabel=Relative program complexity,
    ylabel style={yshift=0mm},
    xtick={1,2,3},
    xticklabels={sliced,accelerated,abstracted},
    extra x ticks={1,2,3},
    extra x tick labels={\sliced,\accelerated,\abstracted},
    extra x tick style={
      xticklabel style={anchor=north,yshift=-1em}
    },
    every node near coord/.append style={font=\scriptsize,/pgf/number format/fixed}
    ]
    \addplot+[
    boxplot prepared={
      median=0.841,
      average=0.776,
      upper quartile=0.998,
      lower quartile=0.683,
      upper whisker=1.0,
      lower whisker=0.185
    },
    ] coordinates {};
    \addplot+[
    boxplot prepared={
      median=0.654,
      average=0.626,
      upper quartile=0.998,
      lower quartile=0.136,
      upper whisker=1.0,
      lower whisker=0.1
    },
    ] coordinates {};
    \addplot+[
    boxplot prepared={
      median=0.215,
      average=0.215,
      upper quartile=0.371,
      lower quartile=0.066,
      upper whisker=0.474,
      lower whisker=0.004
    },
    ] coordinates {}; 
  \end{axis}



\end{tikzpicture}
  \caption{Box plot showing reduction of program complexity of after different source transformation stages, relative to original program. Whiskers are denoting the mix/max values, diamonds the average value}
  \label{fig:speedup}
\end{figure}
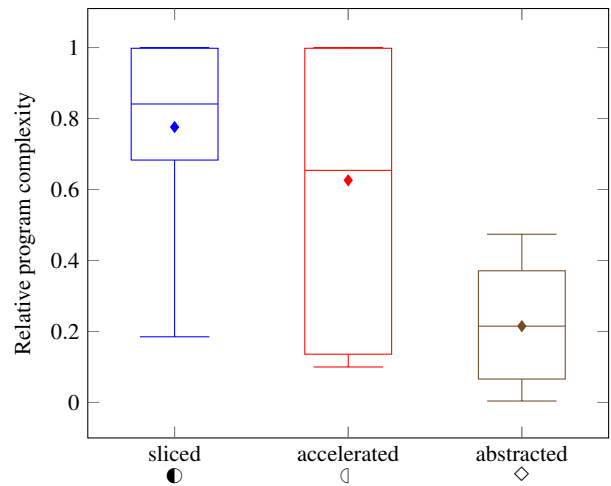

The overall impact of source transformations is quantified in
Figure~\ref{fig:speedup}, where the program size after the respective processing stage is compared to
the instrumented program and over all benchmarks. It can be seen, that
on average each additional processing stage reduces the program
complexity; in average we reach 78\% of the original complexity after
slicing, 63\% after acceleration, and 22\% after
abstraction. Furthermore, it can be seen that in the worst case
slicing and acceleration have no effect, whereas abstraction has a
much more consistent impact on the complexity.

Note that the numbers in Figure~\ref{fig:speedup} exclude those
benchmarks where a timeout had occurred for the instrumented version,
i.e., the original program without our source processing. Here, the
program size could not be determined. For each of those benchmarks, we
have additionally spent 24 hours of processing to determine the
program size, but without success. 

As explained in Section~\ref{sec:wcet}, further reduction of the
analysis time can be reached if a lower bound for the WCET is already
known. This is often the case for safety-critical systems, where
run-time measurements (like high watermarks) are common practice.  The
user would initialize the search algorithm with an observed WCET,
thereby reducing the analysis time. However, this does not change the
structure of the model under analysis, and therefore is not considered
a complexity reduction.

\subsection{Safety and Usability\label{sec:usability-safety}}
Most programs can be analyzed automatically, without any user
input. However, two programs, namely \emph{bs} and \emph{insertsort},
could not be bounded automatically. The loop bounds could not be
deduced by CBMC, which shows up as an infinite loop unwinding
phase. In such cases, the user has to specify an \emph{unwinding
  depth}, at which the loop unwinding stops. CBMC then places an
assertion at this point, which proves that the given depth is
sufficient. In case of an insufficient bound, the model checker
identifies this as a failed property. In case of a too generous bound,
the unwinding may take longer, but WCET is not
overestimated. Therefore, unsound or too imprecise flow constraints do
not refute (or even worsen) the WCET estimate, which makes our approach safer than
others.

Another contribution to the safety of our tool\-chain are the run-time
checks introduced during WCET replay. As discussed earlier, a wide
range of problems can be discovered by this, including confounding of
input data and failure to specify the details of the target
architecture.

Regarding the usability of our toolchain, we argue that the results of
a source-level WCET analysis, especially the possibility to replay the
worst-case path, should increase the level of acceptance for WCET
analysis in a practical setting. As opposed to machine code or
assembly, developers can be expected to be proficient in understanding
a source code, for its higher level of abstraction and higher
readability. By offering the possibility to replay the WCET path
through an ordinarily looking debugging session -- which can be
controlled in the usual ways of stepping, inspecting and resuming --
it becomes intuitive to walk through the path an identify the drivers
of WCET. This is also supported by generating a profiler-like timing
profile of the worst-case path, which is another, well-known view at a
program, and a quick entry point for WCET inspection. A developer can
identify those parts of the WCET path that need an in-depth inspection,
and subsequently dive into debugging.

In summary, our approach is inherently safer than existing approaches
for its protection against unsound user input, and it ensures that
developers need no additional tools or training to analyze and
understand the WCET path of a program.

\subsection{Correctness}
The WCET estimate computed by \tick\ will always be an upper bound of
the actual WCET. This is easy to see as in each step in
\tick, we either over-approximate or preserve the timing
information of machine instructions. In step 1, while the execution time computed for each
basic block in \tick\ (Section \ref{sec:step-1:-backtr}) is
precise, the back-annotation done in Section \ref{sec:back-annotate}
over-approximates the time in the case where multiple basic
blocks map to one source line.  Therefore, at the end of this step,
the WCET of the machine instructions is either preserved or over-approximated in the
source.

The slicing removes only those statements that do not
impact the values of the counter variable, thus preserving the WCET as
per the instrumented program. 

In the next stage, acceleration preserves the computation of the counter variable
and abstraction potentially over-approximates it. So, at the end of
this step, too, WCET is either preserved or
over-approximated. Finally, the iterative search procedure of
Algorithm \ref {wcet-algo} terminates only upon finding an upper bound
of the WCET as per the accelerated or abstracted program. Thus, if
\tick\ scales for the input program, it will provide a safe upper
bound for the WCET.

\subsection{Threats to Validity}
\paragraph{Microarchitectural Features.} We presented our results for a simple
microcontroller. Modern processors, such as ARM SoCs, have
architectural features like caches and pipelines, and memory-mapped
I/O. As shown in \cite{Chattopadhyay2013}, some of these features can be
modeled in C in the form of appropriate constraints. However, adding
such constraints would increase the complexity of the C code. Thus,
while \tick\ can be extended to handle for these features, the
additional constraints may hinder its scalability. In such a
situation, we can eliminate infeasible paths using Model Checking (as
in~\cite{Chattopadhyay2013}) and reduce the search space of the model
checker (as in~\cite{Henry:2014}) while applying our technique.
Furthermore, when the instruction timing depends on the value of operands,
a register value analysis becomes necessary. For example, to determine the
timing of a load instruction, the specific address decides whether this
is cached, or it becomes a slow bus access, possibly with waiting states, or a fast access
to a core-coupled memory. At the moment, we have not addressed how to
carry over such a value analysis to the source code level.

\paragraph{Back-Annotation of Timing.} The mapping of temporal
information from machine instructions to source-level is a rather
technical problem. A compiler could be build which enables complete
traceability in at least one direction. Plans for such a compiler have been
made in~\cite{DBLP:conf/isorc/PuschnerPHKHG13}, unfortunately, we are not
aware of any implementation, which leaves us with the task of matching
instructions to source code. Without any mapping information, this
poses an optimal inexact graph matching problem. Specifically, when
compiler optimization is off, then we expect this to boil down to a
graph minor test (where the edit sequence represents our wanted
mapping and the source graph is fixed), known to have a
polynomial complexity in the number of nodes~\cite{ROBERTSON199565}. With
optimization on, however, it is unclear how the matching could be
established in general. We therefore have to rely on some mapping
information from the compilation process (as given by \emph{addr2line}
in our tool\-chain), and apply techniques to complete the  mapping.

Our mapping strategy assumes that the program is compiled using gcc
4.8.1 for the 8bit AVR microprocessor family~\cite{AVRdatabook}, with
standard compilation options. If we change the target-compiler pair or
the compilation options, then our backmapping strategy may not
work. Since it was established on a case-by-case basis, our strategy
might not be complete, but it was extensive enough to cover all cases
in the M\"alardalen benchmarks.  In general, compiler transformations
are a common problem for all WCET techniques. And, to the best of our
knowledge, there has been no generic solution to this problem, except
to provide support for each architecture, compiler and transformation
individually, often in the form of pattern
matching~\cite{WilhelmWCET08}.

\paragraph{Compiler-Inserted Code.} The compiler may insert
low-level procedures, such as loops for arithmetic shifts or code for
soft-floating point operations. These are not covered in our current
tool. We believe that this is only a matter of tooling. If the source
code for such computations is not available, then \tick\ requires a
pre-computed library of the WCETs of low-level functions, to
facilitate the backmapping for the low-level loops.



\section{Related Work}
\paragraph{WCET Analysis.} An excellent survey about techniques and
tools for WCET analysis was published by Wilhelm et
al.~\cite{WilhelmWCET08}. We refer the reader to this article for a
profound overview on the topic. A commonly used set of benchmark
programs for WCET analysis are the M\"alardalen WCET
benchmarks~\cite{Gustafsson:WCET2010:Benchmarks}.

\paragraph{Model Checking for WCET Analysis.}
There have been studies highlighting the
inadequacy~\cite{wilhelm-no-mc,Lv2008} of Model Checking for the task
of WCET computation, concluding that is does
not scale for WCET estimation. Experiments in \cite{Lv2008} confirmed
that model checkers often face run-time- or memory-infeasibility
for complex programs, whereas the ILP technique can compute the WCET
almost instantly. However, because they used the same benchmarks that
we are using here (on a very similar processor) \emph{and} we come to
the opposite conclusion, this confirms that our shift of the analysis to
the source code indeed mitigates the scalability issue.

Further, there have been instances where a model checker was used as a subanalysis
or optimization step in WCET estimation. Chattopadhyay
et al.~\cite{Chattopadhyay2013} propose the use of AI for cache
analysis and Model Checking for pruning infeasible paths considered by
AI. Marref et al.~\cite{Marref2011} show automatic derivation of exact
program flow constraints for WCET computation using Model Checking for
hypothesis validation. Another proponent for Model Checking in WCET
analysis is found in Metzner~\cite{Metzner2004}, who has shown that it
can improve cache analysis, while not suffering from numerical
instabilities, unlike ILP. However, none of these approaches addressed the
scalability issue of Model Checking.

\paragraph{WCET Analysis at Higher Levels of Abstraction.} 
One of the first works proposing WCET analysis at source level was
published by Puschner \cite{Puschner89}. He introduced the notion of
\emph{timing schemata}, where each source-level construct is assigned
a constant execution time. Many constraints have been imposed on the
programming language, as well as annotation constructs to aid
analysis. A similar approach was described shortly after that in
\cite{Park91}. In fact, the processors they used were very comparable to
the one used in our experiments. However, in their case, all loops
must be statically bounded, and overestimation is a direct result
of assigning a constant execution time to source level constructs
(since the compiler may translate the same construct very differently,
depending on the specific variable types and values).

Holsti~\cite{Holsti2008} proposed modeling execution time as a global
variable, and to use a dependency and value analyses (Presburger
Arithmetic) to determine the value of the variable at the end of a
function and meanwhile exclude infeasible paths. Similar to our
approach, slicing and loop acceleration were suggested. However, he
showed by example that only some infeasible paths can be excluded by
his method, whereas we can precisely detect all infeasible
paths. Furthermore, his approach seems to have scalability issues,
which unfortunately are not detailed further due to a lacking set of
experiments.
Kim et al.~\cite{Kim2009} experimented with computing WCET using Model
Checking at source level on small and simple programs, but without
addressing the scalability issues, nor providing experimental data.

Puschner~\cite{Puschner1998} later computed WCET on an AST-like
representation of a program, where flow constraints are expected to be
given by the user, and assuming that the compiler performs the
back-mapping of actual execution times to the source code. He used ILP to compute the longest path. It should be noted that
any ILP-based approach cannot handle variable execution times of basic
blocks without large overapproximation, whereas Model Checking can
encode such properties with non\-/determinism and range
constraints. Furthermore, complete path reconstruction is not possible either
with that approach.

WCET analysis has also been proposed at an intermediate level in
between source code and machine code, similarly because of easier
analysis of data and control flow.  Altenbernd~\cite{Altenbernd2016}
et al. developed an approximation of WCET which works on an ALF
representation of the program, without analyzing the executable. They
automatically identified a timing model of the intermediate instruction
set through execution and measurement of training
programs. As an effect, the analysis is very efficient, but the result is a possibly
unsafe WCET estimate.


\paragraph{Program Slicing, Acceleration and Abstraction.}
Hatcliff~\cite{Hatcliff:2000} was the first to suggest the use of
program slicing to help scale up Model Checking. In this work, we have
build on the acceleration and abstraction capabilities of LABMC
\cite{labmc}. Different abstractions for improving the precision of
WCET computation or determining loop bounds have been explored by
other researchers. Ermedahl et al.~\cite{Ermedahl2005} show precision
improvement in WCET computations by clustering basic blocks in a
program.  Knoop et al.~\cite{Knoop2012} use recurrence relations to
determine loop bounds in programs. Blazy et al.~\cite{Blazy2014} use
program slicing and loop bound calculation techniques to formally
verify computed loop bounds. \v{C}ern{\'y} et
al.~\cite{Zwirchmayr2014} apply a segment abstraction technique to
improve the precision of WCET computations.  While in these
abstractions could be used in some situations, in general they are
either too restrictive because they do not work for a large class of
programs, or they fail to address the scalability issue arising in the
context of using a model checker to compute the WCET.  Al-Bataineh et
al.~\cite{Al-Bataineh2015} use a precise acceleration of
timed-automata models (with cyclic behavior) for WCET computation, to
scale up their IPET technique. However, these ideas are not readily
applicable to loop acceleration in C programs in the absence of
suitable abstractions.

\paragraph{Timing Debugging.} Understanding how the worst-case timing
is produced is very important for practitioners, but there is only a
small body of work on this topic.
Reconstructing the inputs leading to WCET has been done before by
Ermedahl~\cite{Ermedahl2009}, and is perhaps the closest work in
respect to the degree of detail that is provided on the WCET path. Our
approach, however, uses entirely different methods. While Ermedahl
applies a mixture of static analysis and measurements to perform a
systematic search over the value space with WCET estimation, and only
provides the inputs to the WCET path, our approach is leveraging the
output of the model checker that witnesses the WCET estimate, performs
only a single run of the application, and reconstructs the WCET
inputs, as well as the precise path being taken together with a timing
profile. It is thus less expensive and better integrated with the
actual WCET estimation.

Making the worst-case (and best-case) timing visible in the source
code is a rather old idea, first appeared around
1995~\cite{DBLP:conf/rtas/KoHRAWH96}.  The authors highlighted
WCET/BCET paths in the source code, allowing the user to visually see
the WCET path and how time passes on that path. This work was later
extended in~\cite{DBLP:conf/rtas/ZhaoKWHMU04} to apply genetic
algorithms for minimizing WCET, and still later to introduce compiler
transformations to reduce the WCET (this is, however, beyond the scope
of this article).  A similar, but interactive tool was presented in
\cite{DBLP:conf/rtcsa/HarmonK07}. This was also a tree-based approach
for it was focusing on analysis speed instead of precision.  All these
approaches did not provide tight results, as they were based on timing
trees which did not consider data dependencies; they can only reconstruct the
WCET path w.r.t. locations and time spent there, but lack a specific
trace including variable values. Furthermore, trivial loop bounds had
to be given manually by the user.

The commercial WCET tool RapiTime~\cite{bernat2007identifying}
estimates the timing of software, and enables the user to identify
timing bottlenecks at source-code level. For that, the tool provides a
simple path highlighting, but also allows predicting what would happen
if a specific function would be optimized. However, the tool is
measurement-based and thus cannot give any guarantees. The commercial
tool AbsInt~\cite{ferdinand2008combining} provides time debugging
capabilities at the assembly level, e.g., a call graph and a timing
profile. It is also capable of mapping back this timing information to
a high-level model, where the building blocks of the model are
annotated with their contribution to the WCET. Further, the tool
allows to make changes to the model, and compare the results with the
previous ones, essentially enabling a trial-and-error approach to
improve the WCET. Finally, a similar toolset that realizes a generic
formal interface between a modeling tool and a timing analysis tool
has been presented in~\cite{DBLP:conf/rtns/FuhrmannBHS16}, but relies
on external tools for the actual analysis. All of these tools enable
timing debugging to some extent, but they cannot provide a specific
trace with values, or even allow the user to interactively step
through the same. In contrast, our WCET path has the maximum level of
detail, and can be inspected in a well-known debugger environment,
thus offers deeper and more intuitive explanation of how the WCET path
came to be.

\paragraph{Possible Enhancements.}
One approach that could be combined with ours to further speed up
Model Checking, is that of Henry et al.~\cite{Henry:2014}. They also
employ an SMT solver to compute the WCET (just like our back-end), but
they propose additional constraints to have the solver ignore
infeasible paths. This helps to further increase the scalability, but
under the assumption that the programs are loop-free, or that loops
have been unrolled. This is therefore an enhancement that fits well
our approach.
Brandner~\cite{DBLP:conf/rtns/BrandnerHJ12} computed time criticality
information on all parts of a program, to help focus on optimizing
paths that are close to WCET, and not only those on it. However, this
work only provides a relative ranking of all code blocks (not absolute
numbers on time consumption), and requires an external WCET analyzer
that annotates code blocks with individual WCETs. It can therefore be
viewed as another possible extension for our work.

\section{Concluding Remarks}
We have shown that Model Checking can be a competitive approach to
Worst-Case Execution Time Analysis, in particular when an analysis-friendly
processor is used. The estimates are comparable to and sometimes even
more
precise 
than estimates of the widely-used ILP technique, but with several
practical advantages. Additional to a precise estimate, we also
reconstruct and replay the WCET path, while providing profiling data
and a well-known debugger interface, allowing the user to inspect
arbitrary details on the WCET path at both source code and machine
code level.  Although our approach does not entirely remove the need
for manual inputs from the user, WCET analysis is no longer prone to
human error coming from there, because the model checker also verifies
whether such inputs are sound. If too small bounds are given, an error
is flagged. Too large bounds, on the other hand, only influence the
analysis time, but not the outcome. 
In summary, we therefore arrive at a safer WCET analysis and a more intuitive
understanding of the outcome.

An essential part of our approach is the shift of the analysis from
machine instructions to the source code. Through this, data and
control flows can be tracked more precisely, and source code
transformation techniques can be applied to summarize loops, to remove
statements not related to timing, and to over-approximate larger
programs. As a result, the analysis time can be reduced significantly,
making Model Checking a viable approach to the Worst-Case
Execution Time problem.

What we have shown in this article is merely the first step of
reintroducing Model Checking to Worst-Case Execution Time
Analysis. However, there is substantially more work to be done to
catch up with the proven approaches of combining AI and ILP: Here, we
assumed processors with almost constant instruction
timing. While certainly processors for real-time applications should
be simplified to address the self-made and acknowledged problem of
processors becoming more and more unpredictable, the approach
presented here must be extended to meet current processors half
way. Specifically, we plan to include a register-level value
analysis to be able to handle variable timing due to memory-mapped
I/O, and to propose a specific source-level model for scratchpad
memories. This will result in support for a much wider range of
processors. The challenge in these extensions will mainly be how to
lift models for these architectural behaviors to the source code,
without impairing the scalability of Model Checking too much.
Nevertheless, pursuing this route should be worth the efforts that are
on the way, by reason of the practical benefits over existing
approaches, such as higher automation and usability.

\begin{itemize}
\item 
\end{itemize}


%
\bibliographystyle{spmpsci}
\vspace*{-2em}
\bibliography{literature}

\begin{thebibliography}{10}
\providecommand{\url}[1]{{#1}}
\providecommand{\urlprefix}{URL }
\expandafter\ifx\csname urlstyle\endcsname\relax
  \providecommand{\doi}[1]{DOI~\discretionary{}{}{}#1}\else
  \providecommand{\doi}{DOI~\discretionary{}{}{}\begingroup
  \urlstyle{rm}\Url}\fi

\bibitem{Al-Bataineh2015}
Al-Bataineh, O., Reynolds, M., French, T.: Accelerating worst case execution
  time analysis of timed automata models with cyclic behaviour.
\newblock Formal Aspects of Computing \textbf{27}(5), 917--949 (2015)

\bibitem{Altenbernd2016}
Altenbernd, P., Gustafsson, J., Lisper, B., Stappert, F.: Early execution
  time-estimation through automatically generated timing models.
\newblock Real-Time Systems \textbf{52}(6), 731--760 (2016)

\bibitem{DBLP:journals/tecs/AxerEFGGGJMRRSHW014}
Axer, P., Ernst, R., Falk, H., Girault, A., Grund, D., Guan, N., Jonsson, B.,
  Marwedel, P., Reineke, J., Rochange, C., Sebastian, M., von Hanxleden, R.,
  Wilhelm, R., Yi, W.: Building timing predictable embedded systems.
\newblock {ACM} Trans. Embedded Comput. Syst. \textbf{13}(4), 82:1--82:37
  (2014)

\bibitem{SAFECOMP15}
Becker, M., Neumair, M., S{\"o}hn, A., Chakraborty, S.: {Approaches for
  Software Verification of an Emergency Recovery System for Micro Air
  Vehicles}.
\newblock In: F.~Koornneef, C.~van Gulijk (eds.) Proc.\ Computer Safety,
  Reliability, and Security - 34th International Conference (SAFECOMP),
  \emph{Lecture Notes in Computer Science}, vol. 9337. Springer (2015)

\bibitem{bernat2007identifying}
Bernat, G., Davis, R., Merriam, N., Tuffen, J., Gardner, A., Bennett, M.,
  Armstrong, D.: Identifying opportunities for worst-case execution time
  reduction in an avionics system.
\newblock Ada User Journal \textbf{28}(3), 189--195 (2007)

\bibitem{SVCOMP}
Beyer, D.: Status report on software verification - (competition summary
  {SV-COMP} 2014).
\newblock In: E.~{\'{A}}brah{\'{a}}m, K.~Havelund (eds.) Proc.\ 20th
  International Conference on Tools and Algorithms for the Construction and
  Analysis of Systems ({TACAS}), \emph{Lecture Notes in Computer Science}, vol.
  8413, pp. 373--388. Springer (2014)

\bibitem{Blazy2014}
Blazy, S., Maroneze, A.O., Pichardie, D.: Formal verification of loop bound
  estimation for {WCET} analysis.
\newblock In: E.~Cohen, A.~Rybalchenko (eds.) Proc.\ 5th International
  Conference on Verified Software: Theories, Tools, Experiments ({VSTTE}),
  \emph{Lecture Notes in Computer Science}, vol. 8164, pp. 281--303. Springer
  (2014)

\bibitem{DBLP:conf/rtns/BrandnerHJ12}
Brandner, F., Hepp, S., Jordan, A.: Static profiling of the worst-case in
  real-time programs.
\newblock In: L.~Cucu{-}Grosjean, N.~Navet, C.~Rochange, J.H. Anderson (eds.)
  Proc.\ 20th International Conference on Real-Time and Network Systems
  {(RTNS)}, pp. 101--110. {ACM} (2012)

\bibitem{brumley2011bap}
Brumley, D., Jager, I., Avgerinos, T., Schwartz, E.J.: {BAP:} {A} binary
  analysis platform.
\newblock In: G.~Gopalakrishnan, S.~Qadeer (eds.) Proc.\ 23rd International
  Conference on Computer Aided Verification ({CAV}), \emph{Lecture Notes in
  Computer Science}, vol. 6806, pp. 463--469. Springer (2011)

\bibitem{Zwirchmayr2014}
Cern{\'{y}}, P., Henzinger, T.A., Kov{\'{a}}cs, L., Radhakrishna, A.,
  Zwirchmayr, J.: Segment abstraction for worst-case execution time analysis.
\newblock In: J.~Vitek (ed.) Proc.\ 24th European Symposium on Programming
  Languages and Systems ({ESOP}), \emph{Lecture Notes in Computer Science},
  vol. 9032, pp. 105--131. Springer (2015)

\bibitem{Chattopadhyay2013}
Chattopadhyay, S., Roychoudhury, A.: Scalable and precise refinement of cache
  timing analysis via path-sensitive verification.
\newblock Real-Time Systems \textbf{49}(4), 517--562 (2013)

\bibitem{Clarke2004}
Clarke, E.M., Kroening, D., Lerda, F.: A tool for checking {ANSI-C} programs.
\newblock In: K.~Jensen, A.~Podelski (eds.) Proc.\ 10th International
  Conference on Tools and Algorithms for the Construction and Analysis of
  Systems ({TACAS}), \emph{Lecture Notes in Computer Science}, vol. 2988, pp.
  168--176. Springer (2004)

\bibitem{Clarke:2000}
Clarke Jr., E.M., Grumberg, O., Peled, D.A.: Model Checking.
\newblock MIT Press, Cambridge, MA, USA (1999)

\bibitem{labmc}
Darke, P., Chimdyalwar, B., Venkatesh, R., Shrotri, U., Metta, R.:
  Over-approximating loops to prove properties using bounded model checking.
\newblock In: W.~Nebel, D.~Atienza (eds.) Proc.\ Design, Automation {\&} Test
  in Europe Conference {\&} Exhibition ({DATE}), pp. 1407--1412. {ACM} (2015)

\bibitem{Demyanova2015}
Demyanova, Y., Pani, T., Veith, H., Zuleger, F.: Empirical software metrics for
  benchmarking of verification tools.
\newblock In: D.~Kroening, C.S. Pasareanu (eds.) Proc.\ 27th International
  Conference on Computer Aided Verification ({CAV}), \emph{Lecture Notes in
  Computer Science}, vol. 9206, pp. 561--579. Springer (2015)

\bibitem{Ding2004}
Ding, H., Liang, Y., Mitra, T.: Wcet-centric partial instruction cache locking.
\newblock In: P.~Groeneveld, D.~Sciuto, S.~Hassoun (eds.) Proc.\ 49th Annual
  Design Automation Conference ({DAC}), pp. 412--420. {ACM} (2012)

\bibitem{DBLP:conf/iccd/EdwardsKLLPS09}
Edwards, S.A., Kim, S., Lee, E.A., Liu, I., Patel, H.D., Schoeberl, M.: A
  disruptive computer design idea: Architectures with repeatable timing.
\newblock In: Proc.\ 27th International Conference on Computer Design ({ICCD}),
  pp. 54--59. {IEEE} Computer Society (2009)

\bibitem{Ermedahl2009}
Ermedahl, A., Fredriksson, J., Gustafsson, J., Altenbernd, P.: Deriving the
  worst-case execution time input values.
\newblock In: Proc.\ 21st Euromicro Conference on Real-Time Systems ({ECRTS}),
  pp. 45--54. {IEEE} Computer Society (2009)

\bibitem{Ermedahl2005}
Ermedahl, A., Stappert, F., Engblom, J.: Clustered worst-case execution-time
  calculation.
\newblock {IEEE} Trans. Computers \textbf{54}(9), 1104--1122 (2005)

\bibitem{ferdinand2008combining}
Ferdinand, C., Heckmann, R., Le~Sergent, T., Lopes, D., Martin, B., Fornari,
  X., Martin, F.: Combining a high-level design tool for safety-critical
  systems with a tool for wcet analysis of executables.
\newblock In: Proc.\ 4th European Congress on Embedded Real Time Software
  ({ERTS}). {SIA}/{AAAF}/{SEE} (2008)

\bibitem{DBLP:conf/rtns/FuhrmannBHS16}
Fuhrmann, I., Broman, D., von Hanxleden, R., Schulz{-}Rosengarten, A.: Time for
  reactive system modeling: Interactive timing analysis with hotspot
  highlighting.
\newblock In: A.~Plantec, F.~Singhoff, S.~Faucou, L.M. Pinho (eds.) Proc.\ 24th
  International Conference on Real-Time Networks and Systems {(RTNS)}, pp.
  289--298. {ACM} (2016)

\bibitem{ARMv7TDMI}
Furber, S.: ARM System-on-Chip Architecture.
\newblock Addison Wesley (2000)

\bibitem{Gulwani2009}
Gulwani, S., Jain, S., Koskinen, E.: Control-flow refinement and progress
  invariants for bound analysis.
\newblock In: M.~Hind, A.~Diwan (eds.) Proc.\ Conference on Programming
  Language Design and Implementation ({PLDI}), pp. 375--385. {ACM} (2009)

\bibitem{Gustafsson:WCET2010:Benchmarks}
Gustafsson, J., Betts, A., Ermedahl, A., Lisper, B.: The m{\"{a}}lardalen
  {WCET} benchmarks: Past, present and future.
\newblock In: B.~Lisper (ed.) Proc.\ 10th International Workshop on Worst-Case
  Execution Time Analysis ({WCET}), \emph{{OASICS}}, vol.~15, pp. 136--146.
  Schloss Dagstuhl - Leibniz-Zentrum fuer Informatik, Germany (2010)

\bibitem{Gustafsson2006}
Gustafsson, J., Ermedahl, A., Sandberg, C., Lisper, B.: Automatic derivation of
  loop bounds and infeasible paths for wcet analysis using abstract execution.
\newblock In: Proc.\ 27th International Real-Time Systems Symposium {(RTSS)},
  pp. 57--66 (2006)

\bibitem{DBLP:conf/rtcsa/HarmonK07}
Harmon, T., Klefstad, R.: Interactive back-annotation of worst-case execution
  time analysis for java microprocessors.
\newblock In: Proc.\ 13th International Conference on Embedded and Real-Time
  Computing Systems and Applications {(RTCSA)}, pp. 209--216. {IEEE} Computer
  Society (2007)

\bibitem{Hatcliff:2000}
Hatcliff, J., Dwyer, M.B., Zheng, H.: Slicing software for model construction.
\newblock Higher-Order and Symbolic Computation \textbf{13}(4), 315--353 (2000)

\bibitem{healy2000supporting}
Healy, C.A., Sj{\"{o}}din, M., Rustagi, V., Whalley, D.B., van Engelen, R.:
  Supporting timing analysis by automatic bounding of loop iterations.
\newblock Real-Time Systems \textbf{18}(2/3), 129--156 (2000)

\bibitem{Henry:2014}
Henry, J., Asavoae, M., Monniaux, D., Maiza, C.: How to compute worst-case
  execution time by optimization modulo theory and a clever encoding of program
  semantics.
\newblock In: Y.~Zhang, P.~Kulkarni (eds.) Proc.\ 15th Conference on Languages,
  Compilers and Tools for Embedded Systems ({LCTES}), pp. 43--52. {ACM} (2014)

\bibitem{Holsti2008}
Holsti, N.: Computing time as a program variable: a way around infeasible
  paths.
\newblock In: R.~Kirner (ed.) Proc.\ 8th Intl. Workshop on Worst-Case Execution
  Time {(WCET)} Analysis, \emph{{OASICS}}, vol.~8. Internationales Begegnungs-
  und Forschungszentrum fuer Informatik (IBFI), Schloss Dagstuhl, Germany
  (2008)

\bibitem{holsti2002status}
Holsti, N., Saarinen, S.: {Status of the Bound-T WCET tool}.
\newblock Space Systems Finland Ltd  (2002)

\bibitem{Kim2009}
Kim, S., Patel, H.D., Edwards, S.A.: {Using a Model Checker to Determine
  Worst-Case Execution Time}.
\newblock Tech. rep., Columbia University (2009).
\newblock CUCS-038-09

\bibitem{DBLP:conf/isorc/KirnerP08}
Kirner, R., Puschner, P.P.: Obstacles in worst-case execution time analysis.
\newblock In: Proc.\ 11th {IEEE} International Symposium on Object-Oriented
  Real-Time Distributed Computing ({ISORC}), pp. 333--339. {IEEE} Computer
  Society (2008)

\bibitem{Knoop2012}
Knoop, J., Kov{\'{a}}cs, L., Zwirchmayr, J.: Symbolic loop bound computation
  for {WCET} analysis.
\newblock In: E.M. Clarke, I.~Virbitskaite, A.~Voronkov (eds.) Proc.\ 8th
  International Conference Perspectives of Systems Informatics ({PSI}), Revised
  Selected Papers, \emph{Lecture Notes in Computer Science}, vol. 7162, pp.
  227--242. Springer (2012)

\bibitem{DBLP:conf/rtas/KoHRAWH96}
Ko, L., Healy, C.A., Ratliff, E., Arnold, R.D., Whalley, D.B., Harmon, M.G.:
  Supporting the specification and analysis of timing constraints.
\newblock In: Proc.\ 2nd Real-Time Technology and Applications Symposium
  ({RTAS}), pp. 170--178. {IEEE} Computer Society (1996)

\bibitem{AVRdatabook}
Kuhnel, C.: AVR RISC Microcontroller Handbook, first edn.
\newblock Newnes (1998)

\bibitem{Kuo2010}
Kuo, M.M.Y., Yoong, L.H., Andalam, S., Roop, P.S.: {Determining the worst-case
  reaction time of IEC 61499 function blocks}.
\newblock In: Proc.\ 8th IEEE International Conference on Industrial
  Informatics, pp. 1104--1109. IEEE (2010)

\bibitem{Li1997}
Li, Y.T., Malik, S.: {Performance analysis of embedded software using implicit
  path enumeration}.
\newblock IEEE Transactions on Computer-Aided Design of Integrated Circuits and
  Systems \textbf{16}(12) (1997)

\bibitem{PRET2008}
Lickly, B., Liu, I., Kim, S., Patel, H.D., Edwards, S.A., Lee, E.A.:
  Predictable programming on a precision timed architecture.
\newblock In: E.R. Altman (ed.) Proc.\ International Conference on Compilers,
  Architecture, and Synthesis for Embedded Systems, ({CASES}), pp. 137--146.
  {ACM} (2008)

\bibitem{Lv2008}
Lv, M., Gu, Z., Guan, N., Deng, Q., Yu, G.: Performance comparison of
  techniques on static path analysis of {WCET}.
\newblock In: C.~Xu, M.~Guo (eds.) Proc.\ International Conference on Embedded
  and Ubiquitous Computing ({EUC}), pp. 104--111. {IEEE} Computer Society
  (2008)

\bibitem{Marref2011}
Marref, A.: Fully-automatic derivation of exact program-flow constraints for a
  tighter worst-case execution-time analysis.
\newblock In: L.~Carro, A.D. Pimentel (eds.) Proc.\ International Conference on
  Embedded Computer Systems: Architectures, Modeling, and Simulation ({SAMOS}),
  pp. 200--208. {IEEE} (2011)

\bibitem{metta2016tic}
Metta, R., Becker, M., Bokil, P., Chakraborty, S., Venkatesh, R.: {TIC:} a
  scalable model checking based approach to {WCET} estimation.
\newblock In: T.~Kuo, D.B. Whalley (eds.) Proc.\ 17th Conference on Languages,
  Compilers, Tools, and Theory for Embedded Systems ({LCTES}), pp. 72--81.
  {ACM} (2016)

\bibitem{Metzner2004}
Metzner, A.: Why model checking can improve {WCET} analysis.
\newblock In: R.~Alur, D.A. Peled (eds.) Proc.\ 16th International Conference
  on Computer Aided Verification ({CAV}), \emph{Lecture Notes in Computer
  Science}, vol. 3114, pp. 334--347. Springer (2004)

\bibitem{mitra_et_al:DR:2017:6953}
Mitra, T., Teich, J., Thiele, L.: {Adaptive Isolation for Predictability and
  Security (Dagstuhl Seminar 16441)}.
\newblock Dagstuhl Reports \textbf{6}(10), 120--153 (2017)

\bibitem{Mittal16}
Mittal, S.: A survey of techniques for cache locking.
\newblock {ACM} Trans. Design Autom. Electr. Syst. \textbf{21}(3), 49:1--49:24
  (2016)

\bibitem{Park91}
Park, C.Y., Shaw, A.C.: Experiments with a program timing tool based on
  source-level timing schema.
\newblock {IEEE} Computer \textbf{24}(5), 48--57 (1991)

\bibitem{Pingali1995}
Pingali, K., Bilardi, G.: {APT:} {A} data structure for optimal control
  dependence computation.
\newblock In: D.W. Wall (ed.) Proc.\ Conference on Programming Language Design
  and Implementation {(PLDI)}, pp. 32--46. {ACM} (1995)

\bibitem{Puschner2002}
Puschner, P.: {Is WCET Analysis a Non-Problem? -- Towards New Software and
  Hardware Architectures}.
\newblock In: G.~Bernat (ed.) Proc.\ 2nd International Workshop on Worst-Case
  Execution Time Analysis ({WCET}), pp. 89--92. Technical University of Vienna,
  Austria (2002)

\bibitem{Puschner1998}
Puschner, P.P.: A tool for high-level language analysis of worst-case execution
  times.
\newblock In: Proc.\ 10th Euromicro Conference on Real-Time Systems {(ECRTS}),
  pp. 130--137. {IEEE} Computer Society (1998)

\bibitem{Puschner89}
Puschner, P.P., Koza, C.: Calculating the maximum execution time of real-time
  programs.
\newblock Real-Time Systems \textbf{1}(2), 159--176 (1989)

\bibitem{DBLP:conf/isorc/PuschnerPHKHG13}
Puschner, P.P., Prokesch, D., Huber, B., Knoop, J., Hepp, S., Gebhard, G.: The
  {T-CREST} approach of compiler and wcet-analysis integration.
\newblock In: Proc.\ 16th International Symposium on
  Object/Component/Service-Oriented Real-Time Distributed Computing, ({ISORC}),
  pp. 1--8. {IEEE} Computer Society (2013)

\bibitem{Raymond2013}
Raymond, P., Maiza, C., Parent{-}Vigouroux, C., Carrier, F.: Timing analysis
  enhancement for synchronous program.
\newblock In: M.~Auguin, R.~de~Simone, R.I. Davis, E.~Grolleau (eds.) Proc.\
  21st International Conference on Real-Time Networks and Systems ({RTNS}), pp.
  141--150. {ACM} (2013)

\bibitem{ROBERTSON199565}
Robertson, N., Seymour, P.: Graph minors .xiii. the disjoint paths problem.
\newblock Journal of Combinatorial Theory, Series B \textbf{63}(1), 65 -- 110
  (1995)

\bibitem{JOP}
Schoeberl, M.: {JOP: A Java Optimized Processor}.
\newblock In: R.~Meersman, Z.~Tari (eds.) Proc.\ International Workshop On The
  Move to Meaningful Internet Systems ({OTM}), pp. 346--359. Springer Berlin
  Heidelberg (2003)

\bibitem{Souyris2005}
Souyris, J., Pavec, E.L., Himbert, G., J\'{e}gu, V., Borios, G., Heckmann, R.:
  {Computing the Worst Case Execution Time of an Avionics Program By Abstract
  Interpretation}.
\newblock In: Proc.\ 5th Intl Workshop on Worst-Case Execution Time ({WCET})
  Analysis (2005)

\bibitem{SPARCv7}
{Sun Microsystems Inc.}: {The SPARC Architecture Manual, Version 7}.
\newblock Sun Microsystems Inc. (1987)

\bibitem{Weiser:1981}
Weiser, M.: Program slicing.
\newblock In: S.~Jeffrey, L.G. Stucki (eds.) Proc.\ 5th International
  Conference on Software Engineering ({ICSE}), pp. 439--449. {IEEE} Computer
  Society (1981)

\bibitem{wilhelm-no-mc}
Wilhelm, R.: Why {AI} + {ILP} is good for {WCET}, but {MC} is not, nor {ILP}
  alone.
\newblock In: B.~Steffen, G.~Levi (eds.) Proc.\ 5th International Conference on
  Verification, Model Checking, and Abstract Interpretation ({VMCAI}),
  \emph{Lecture Notes in Computer Science}, vol. 2937, pp. 309--322. Springer
  (2004)

\bibitem{WilhelmWCET08}
Wilhelm, R., Engblom, J., Ermedahl, A., Holsti, N., Thesing, S., Whalley, D.,
  Bernat, G., Ferdinand, C., Heckmann, R., Mitra, T., Mueller, F., Puaut, I.,
  Puschner, P., Staschulat, J., Stenstr\"{o}m, P.: {The Worst-Case Execution
  Time Problem -- Overview of Methods and Survey of Tools}.
\newblock ACM Trans. Embed. Comput. Syst. \textbf{7}(3), 36:1--36:53 (2008)

\bibitem{DBLP:conf/rtas/ZhaoKWHMU04}
Zhao, W., Kulkarni, P.A., Whalley, D.B., Healy, C.A., Mueller, F., Uh, G.:
  Tuning the {WCET} of embedded applications.
\newblock In: Proc.\ 10th Real-Time and Embedded Technology and Applications
  Symposium {(RTAS)}, pp. 472--481. {IEEE} Computer Society (2004)

\end{thebibliography}

\end{document}